%% file: main.tex
\documentclass[12pt]{article}
\usepackage{amsmath,amssymb,mathtools,bm,geometry}
\usepackage{setspace}
\usepackage{microtype}
\usepackage{needspace}
\usepackage{natbib}
\usepackage[most]{tcolorbox}
\tcbuselibrary{listings,skins,breakable}
\usepackage{xcolor}
\usepackage{graphicx}
\usepackage{subcaption}
\usepackage{booktabs}
\usepackage{lscape}
\usepackage{multirow}
\usepackage{array}
\usepackage{float}
\usepackage{tikz}
\usetikzlibrary{arrows.meta, positioning, fit, backgrounds, calc}
\usepackage{hyperref}
\hypersetup{colorlinks=true,linkcolor=blue!60!black,citecolor=blue!60!black,urlcolor=blue!60!black}

\newcommand{\papertitle}{Power Analysis for Prediction-Powered Inference}

\newcommand{\papersummary}{Modern studies increasingly leverage outcomes predicted by machine learning and artificial intelligence (AI/ML) models, and recent work, such as prediction-powered inference (PPI), has developed valid downstream statistical inference procedures. However, classical power and sample size formulas do not readily account for these predictions. In this work, we tackle a simple yet practical question: given a new AI/ML model with high predictive power, how many labeled samples are needed to achieve a desired level of statistical power? We derive closed-form power formulas by characterizing the asymptotic variance of the PPI estimator and applying Wald test inversion to obtain the required labeled sample size. Our results cover widely used settings including two-sample comparisons and risk measures in $2\times2$ tables. We find that a useful rule of thumb is that the reduction in required labeled samples relative to classical designs scales roughly with the $R^2$ between the predictions and the ground truth. Our analytical formulas are validated using Monte Carlo simulations, and we illustrate the framework in three contemporary biomedical applications spanning single-cell transcriptomics, clinical blood pressure measurement, and dermoscopy imaging. We provide our software as an \texttt{R} package and online calculators at \url{https://github.com/yiqunchen/pppower}.}
\newcommand{\paperkeywords}{Artificial intelligence; Experimental design; Label efficiency; Power analysis; Prediction-powered inference; Sample size.}
\newcommand{\PPICorrespondenceEmail}{yiqun.t.chen@gmail.com}

\newcommand{\PPIAffiliations}{%
\mbox{$^{1}$}Department of Biostatistics, Johns Hopkins University, Baltimore, Maryland 21205, U.S.A.\\
\mbox{$^{2}$}Department of Computer Science, Johns Hopkins University, Baltimore, Maryland 21205, U.S.A.\\
\texttt{\PPICorrespondenceEmail}}

\newcommand{\PPIAuthorBlockArxiv}{%
\parbox{0.9\textwidth}{\centering
Yiqun T. Chen$^{1,2}$, Moran Guo$^{1}$, and Shengyi Li$^{1}$\\[0.4em]
\small \PPIAffiliations}}

\makeatletter
\@ifundefined{theorem}{\newtheorem{theorem}{Theorem}}{}
\@ifundefined{proposition}{\newtheorem{proposition}[theorem]{Proposition}}{}
\@ifundefined{corollary}{\newtheorem{corollary}[theorem]{Corollary}}{}
\@ifundefined{assumption}{\newtheorem{assumption}{Assumption}}{}
\@ifundefined{lemma}{}{}
\@ifundefined{remark}{}{}
\@ifundefined{definition}{}{}
\makeatother

\tcbset{colback=gray!5!white, colframe=gray!50!black, boxrule=0.4pt, arc=2pt,
        left=6pt, right=6pt, top=4pt, bottom=4pt}

\newcommand{\E}{\mathbb{E}}
\newcommand{\Var}{\mathrm{Var}}
\newcommand{\Cov}{\mathrm{Cov}}
\newcommand{\Corr}{\mathrm{Corr}}

\newcommand{\wh}{\widehat}
\newcommand{\wt}{\widetilde}
\newcommand{\yiqun}[1]{}
\newcommand{\shengyi}[1]{}

\providecommand{\PPIIntroFigurePlacement}{}
\providecommand{\PPIMeanValidationFigurePlacement}{}
\providecommand{\PPIBinaryTableFigurePlacement}{}
\providecommand{\PPIRegressionFigurePlacement}{}
\providecommand{\PPIApplicationsFigurePlacement}{}
\providecommand{\PPIFiguresAtEnd}{}
\providecommand{\PPIBibliographyBlock}{}

\newcommand{\PPIIntroFigureBlock}{%
\begin{figure}[!t]
\centering
\begin{subfigure}[t]{0.48\linewidth}
\centering
\begin{minipage}[t][5.4cm][c]{\linewidth}
\centering
\begin{tikzpicture}[
  >=Latex,
  font=\footnotesize,
  box/.style={draw=gray!65, fill=gray!5, rounded corners=2pt, align=center,
              text width=7.0cm, inner sep=4pt},
  accent/.style={draw=orange!80!black, fill=orange!10, rounded corners=2pt,
                 align=center, text width=7.0cm, inner sep=4pt},
  flow/.style={-Latex, thick, draw=gray!55}
]
\node[box] (notation)
  {$Y$: true labels,\ $f$: predictions,\ $n$: paired,\ $N$: predictions only};
\node[box] (est)
  [below=0.45cm of notation]
  {\textbf{Estimator}\quad $\hat{\theta}_{\lambda}=\bar{Y}_{n}+\lambda(\bar{\tilde{f}}_{N}-\bar{f}_{n})$};
\node[accent, below=0.45cm of est] (invert)
  {\textbf{Power and sample size}\\[2pt]
  invert $\Var(\hat{\theta}_{\lambda})$ to obtain required labels $n^\star$};
\draw[flow] (notation) -- (est);
\draw[flow] (est) -- (invert);
\end{tikzpicture}
\end{minipage}
\caption{Estimator-and-inversion schematic.}
\end{subfigure}\hfill
\begin{subfigure}[t]{0.48\linewidth}
\centering
\begin{minipage}[t][5.4cm][c]{\linewidth}
\centering
\includegraphics[width=\linewidth,height=5.4cm,keepaspectratio]{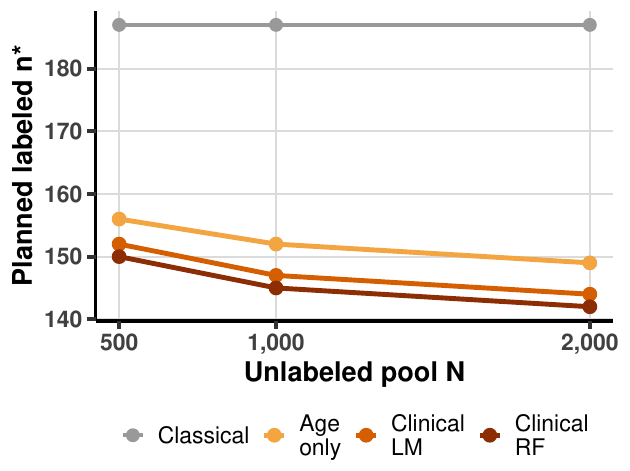}
\end{minipage}
\caption{NHANES planning by surrogate model.}
\end{subfigure}
\caption{Prediction-powered planning. Panel~(a) shows the notation,
estimator, and variance inversion. Panel~(b) shows the NHANES
sample-size plan for $\Delta = 4$~mmHg: gray is the classical design,
and the orange curves are the age-only linear model, the richer
clinical linear model, and the random forest surrogate.}
\label{fig:intro-nhanes}
\end{figure}
}

\newcommand{\PPIMeanValidationFigureBlock}{%
\begin{figure}[!t]
\centering
\begin{subfigure}[b]{0.95\linewidth}
\includegraphics[width=\linewidth]{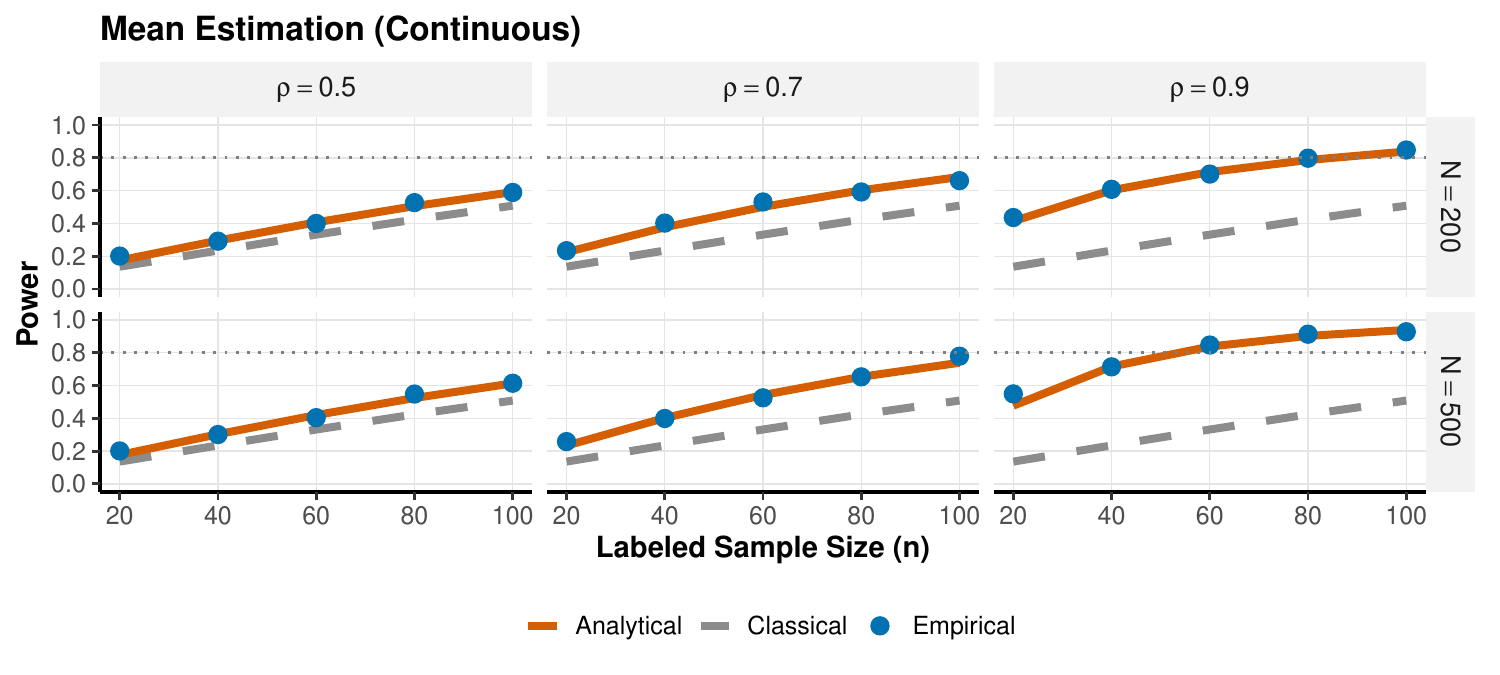}
\caption{Gaussian outcomes.}
\end{subfigure}
\\[6pt]
\begin{subfigure}[b]{0.95\linewidth}
\includegraphics[width=\linewidth]{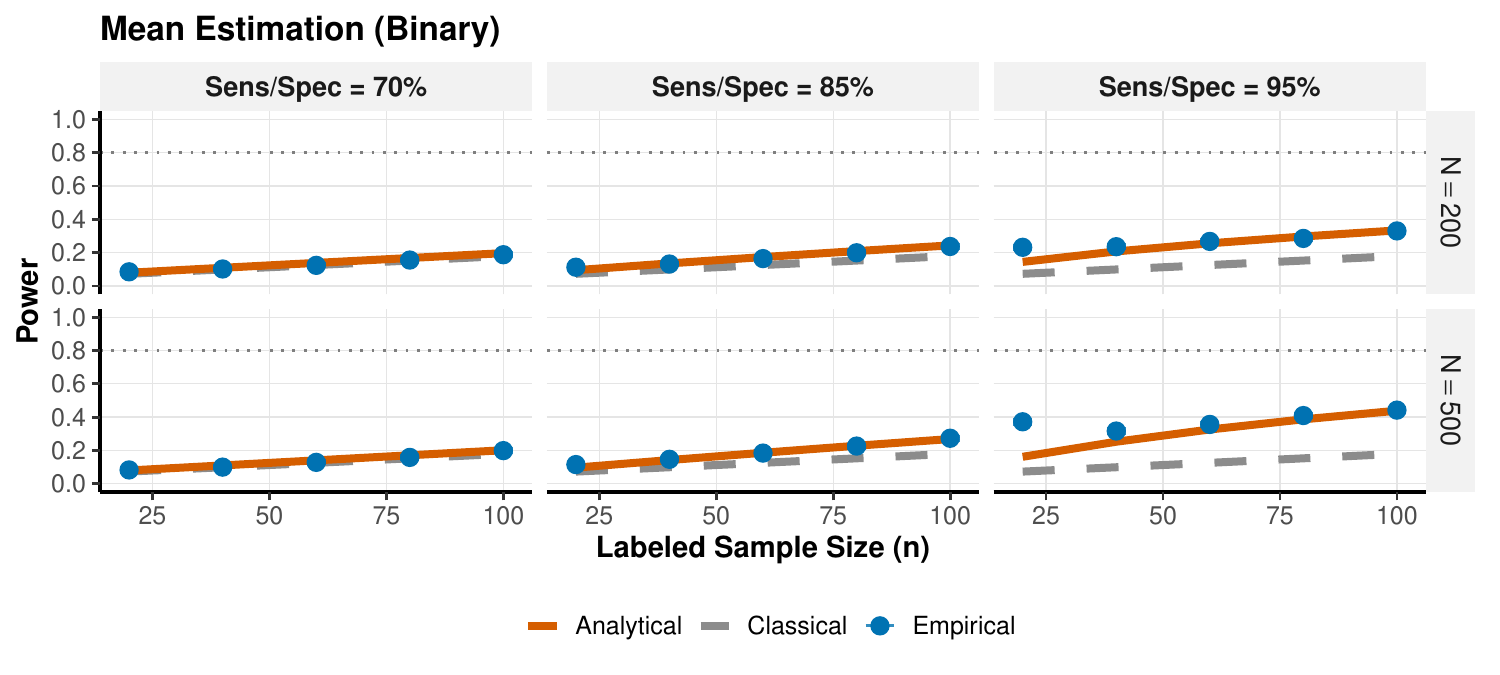}
\caption{Binary outcomes.}
\end{subfigure}
\caption{One-sample mean validation: empirical power (points, with 95\% Monte Carlo
  error bars for the binary setting) versus theoretical power (lines)
  for \texttt{PPI++} with oracle $\lambda^\star$.}
\label{fig:panels-AB}
\end{figure}
}

\newcommand{\PPIBinaryTableFigureBlock}{%
\begin{figure}[!t]
\centering
\includegraphics[width=0.95\linewidth]{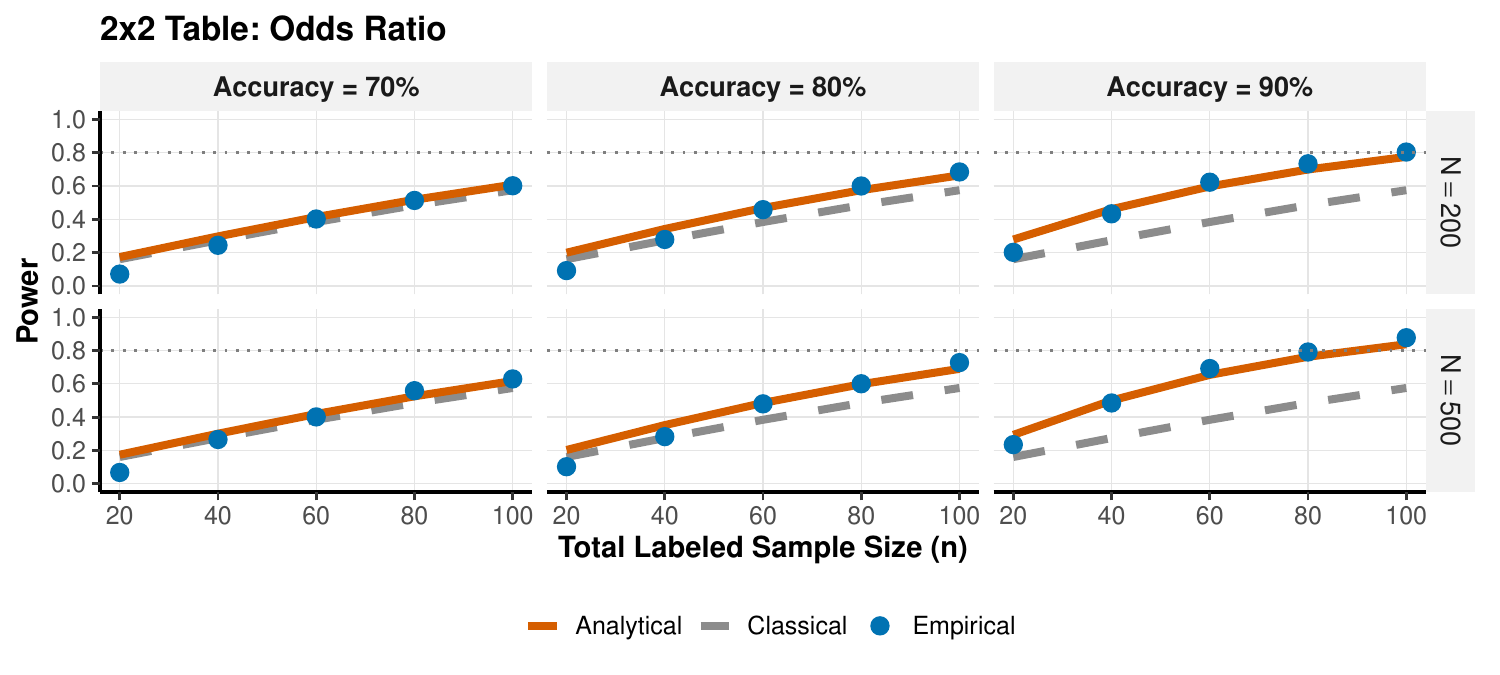}\\[0.4em]
\includegraphics[width=0.95\linewidth]{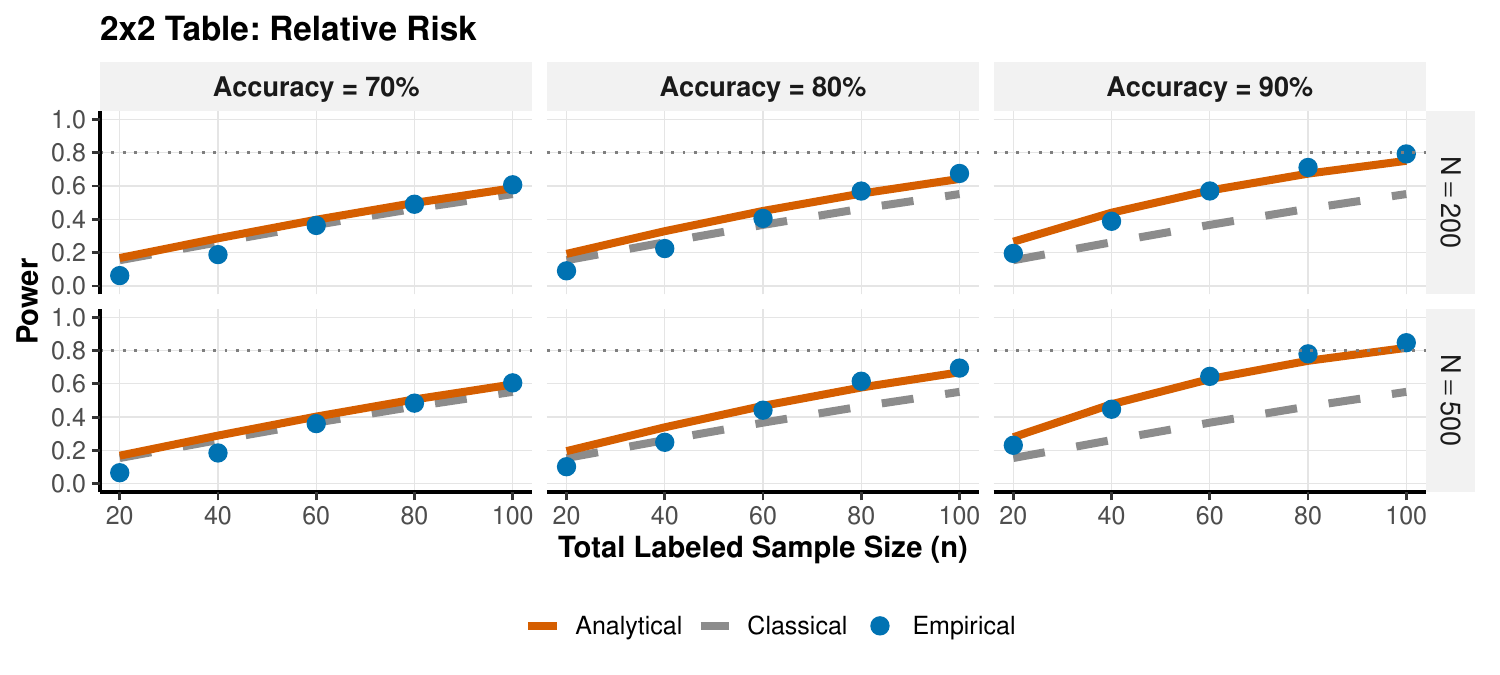}
\caption{Odds-ratio and relative-risk validation in $2\times 2$ tables---analytical (lines)
  versus empirical (points) \texttt{PPI++} power for odds ratio (top) and
  relative risk (bottom). Dashed lines show classical power.}
\label{fig:setting-U}
\end{figure}
}

\newcommand{\PPIRegressionFigureBlock}{%
\begin{figure}[!t]
\centering
\begin{subfigure}[t]{0.95\linewidth}
\centering
\includegraphics[width=\linewidth]{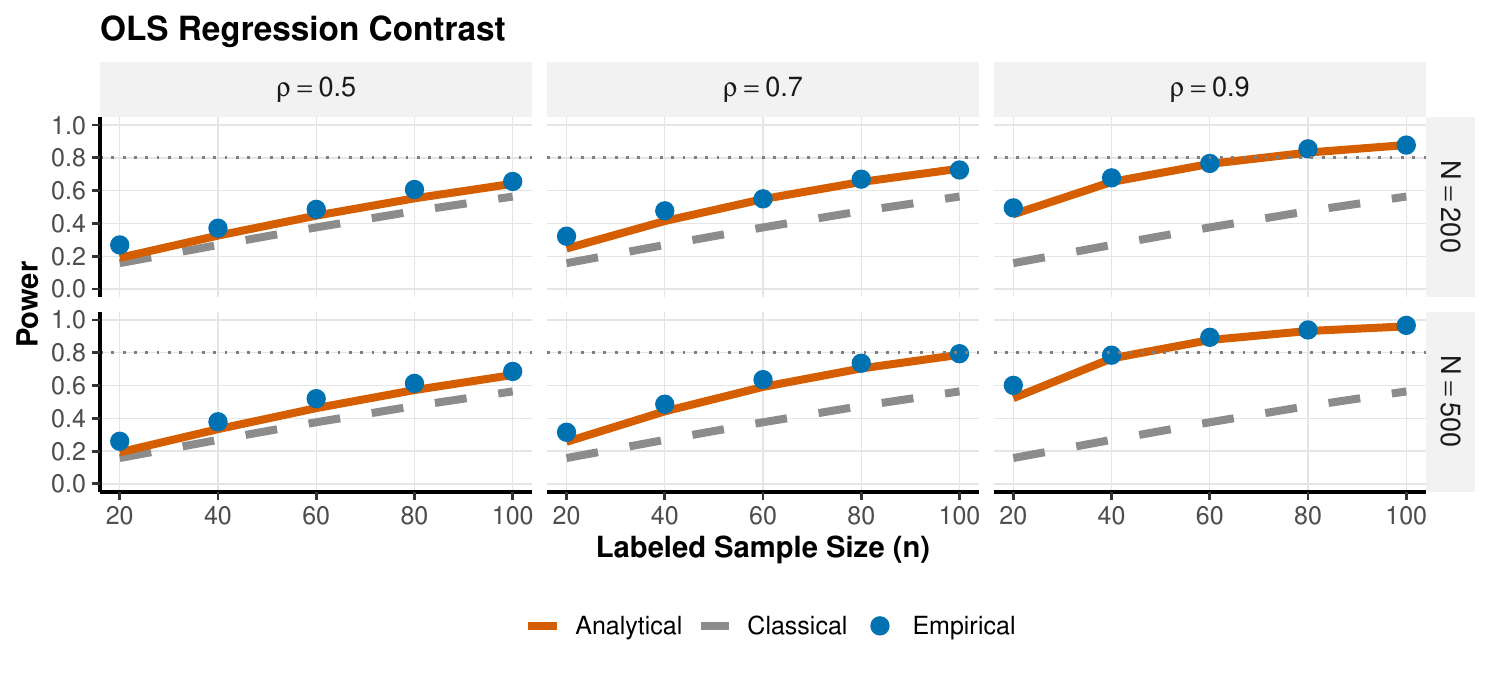}
\caption{OLS beta-coefficient contrast validation.}
\end{subfigure}

\vspace{0.6em}
\begin{subfigure}[t]{0.95\linewidth}
\centering
\includegraphics[width=\linewidth]{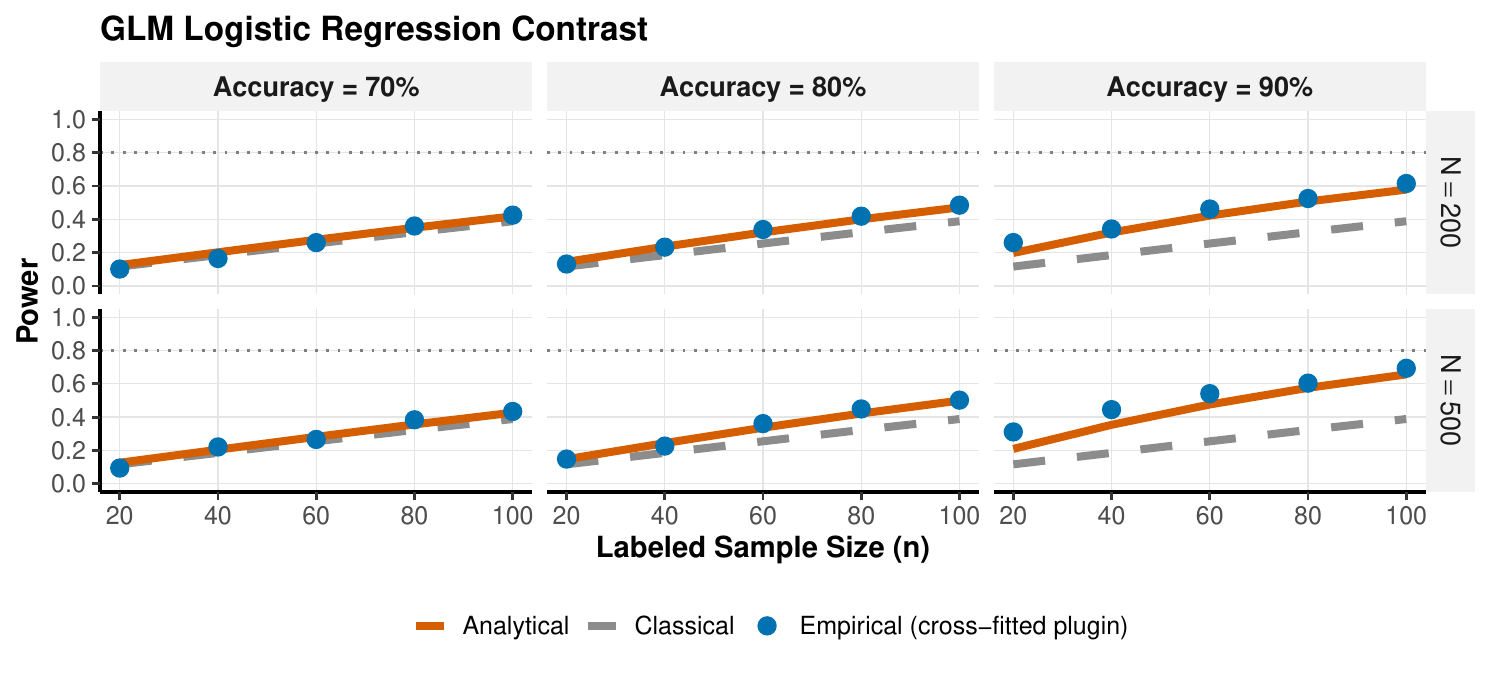}
\caption{Logistic-regression contrast validation.}
\end{subfigure}
\caption{Regression-contrast validation: analytical (lines) versus
  empirical (points) \texttt{PPI++} power. In panel~(b), the empirical
  logistic-regression points use a two-fold cross-fitted plug-in
  $\hat\lambda$ estimated within each replicate, while the analytical
  curve uses the large-reference-sample approximation. Dashed lines show
  the corresponding classical power curves.}
\label{fig:setting-ST}
\end{figure}
}

\newcommand{\PPIApplicationsFigureBlock}{%
\begin{figure}[p]
\centering
\begin{minipage}[t]{0.31\linewidth}
\centering
\textbf{Study overview}
\end{minipage}\hfill
\begin{minipage}[t]{0.31\linewidth}
\centering
\textbf{Planned labels}
\end{minipage}\hfill
\begin{minipage}[t]{0.31\linewidth}
\centering
\textbf{Held-out power}
\end{minipage}

\vspace{0.45em}

\makebox[\linewidth][l]{\small\textbf{Baron scRNA-seq}}\par
\vspace{0.2em}
\includegraphics[width=0.31\linewidth]{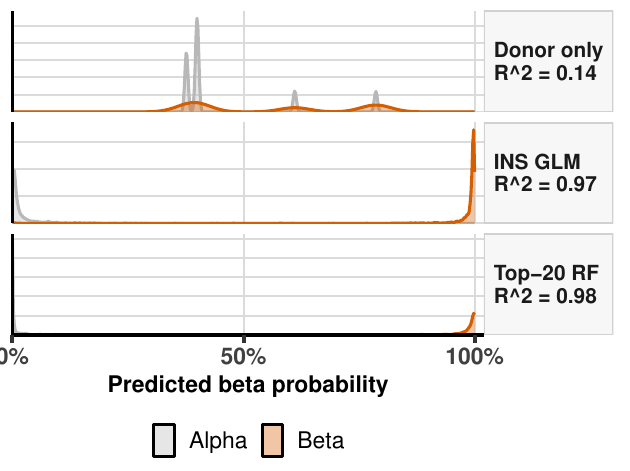}\hfill
\includegraphics[width=0.31\linewidth]{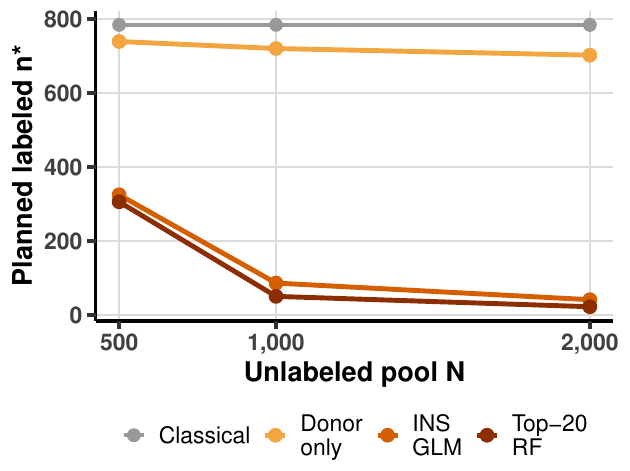}\hfill
\includegraphics[width=0.31\linewidth]{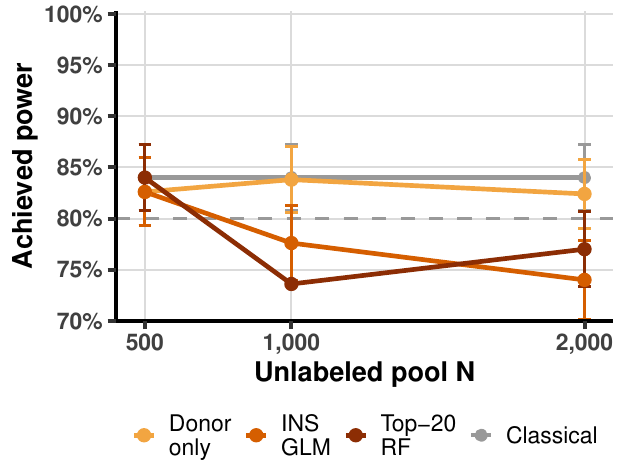}

\vspace{0.55em}

\makebox[\linewidth][l]{\small\textbf{NHANES systolic blood pressure}}\par
\vspace{0.2em}
\includegraphics[width=0.31\linewidth]{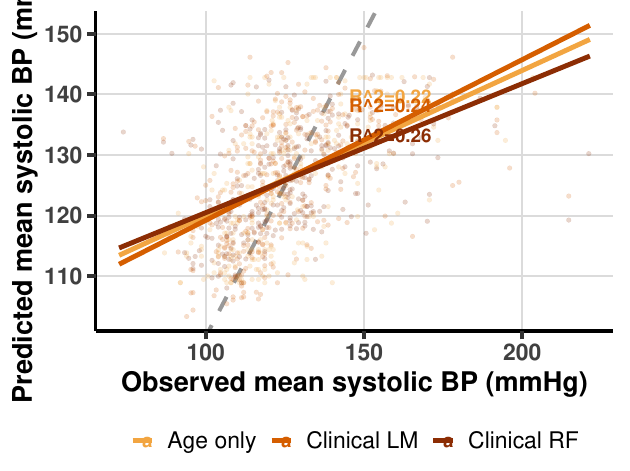}\hfill
\includegraphics[width=0.31\linewidth]{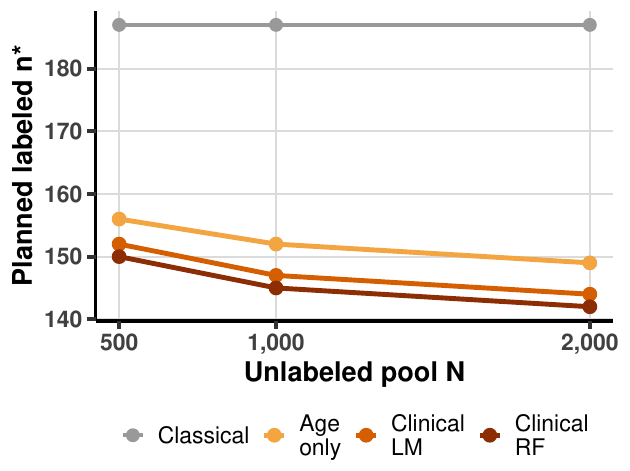}\hfill
\includegraphics[width=0.31\linewidth]{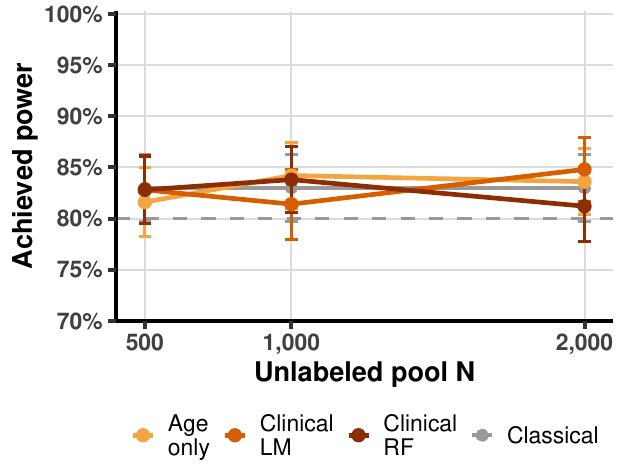}

\vspace{0.55em}

\makebox[\linewidth][l]{\small\textbf{ISIC melanoma}}\par
\vspace{0.2em}
\includegraphics[width=0.31\linewidth]{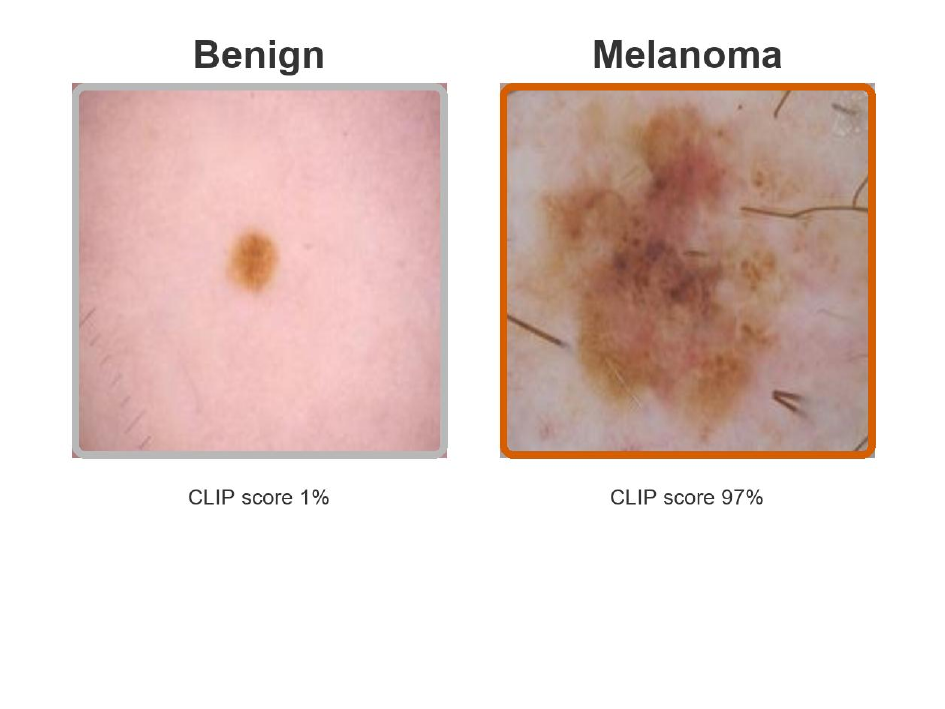}\hfill
\includegraphics[width=0.31\linewidth]{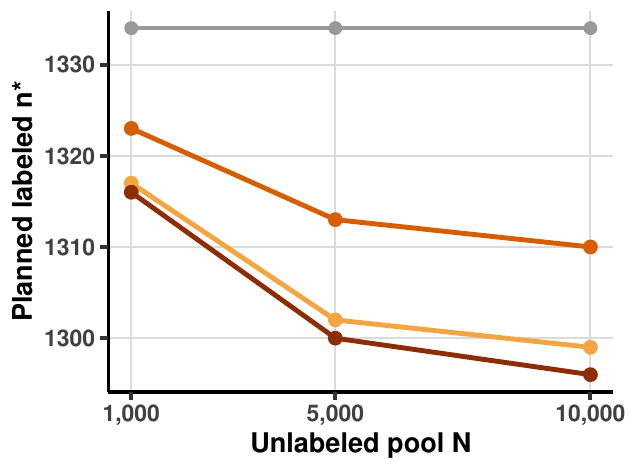}\hfill
\includegraphics[width=0.31\linewidth]{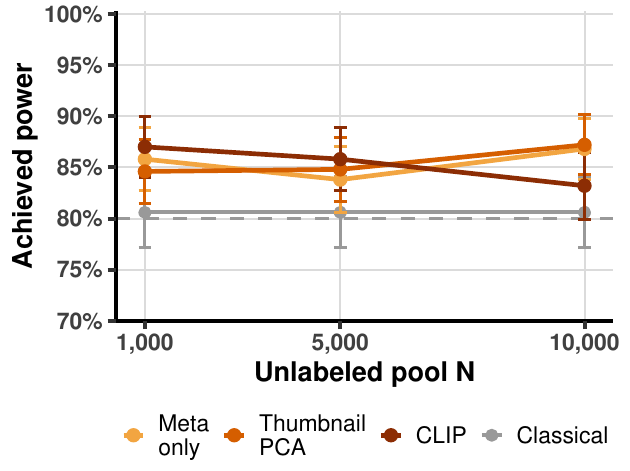}
\caption{Real-data planning and held-out validation across three
  biomedical case studies. Each row corresponds to one application
  (Baron scRNA-seq, NHANES systolic blood pressure, and ISIC
  melanoma), and the three columns give a study overview, the planned
  labeled sample size, and the held-out achieved power.}
\label{fig:applications-summary}
\end{figure}
}

\title{\papertitle}
\author{\PPIAuthorBlockArxiv}
\date{}

\renewcommand{\PPIIntroFigurePlacement}{\PPIIntroFigureBlock}
\renewcommand{\PPIMeanValidationFigurePlacement}{\PPIMeanValidationFigureBlock}
\renewcommand{\PPIBinaryTableFigurePlacement}{\PPIBinaryTableFigureBlock}
\renewcommand{\PPIRegressionFigurePlacement}{\PPIRegressionFigureBlock}
\renewcommand{\PPIApplicationsFigurePlacement}{\PPIApplicationsFigureBlock}
\renewcommand{\PPIBibliographyBlock}{%
{\begingroup
\setlength{\bibsep}{2pt}
\bibliographystyle{plainnat}
\bibliography{references}
\endgroup}
}

\begin{document}
\maketitle
\begin{abstract}
\sloppy
\noindent\papersummary
\end{abstract}
\fussy
\vspace{1em}
\section{Introduction}
\label{sec:intro}

Gold-standard labels are fundamental for scientific discoveries but remain time-consuming and expensive: Annotating cell types in single-cell RNA-seq requires expert curation of marker
genes~\citep{baron2016single}; measuring clinical outcomes in randomized trials demands patient follow-up and laboratory assays~\citep{rodger2012diagnostic}; and evaluating the quality of large language model (LLM) responses to complex questions needs human expert judgments~\citep{chiang2024chatbot}. At the same time, predictions from machine learning (ML) and artificial intelligence (AI) models have become more accurate and available, where millions of cells can be labeled in minutes, prognostic models can score electronic health records at negligible cost, and ``LLM-as-a-judge'' pipelines, where an LLM serves as the human expert, can evaluate millions of responses
in minutes. With the rise of these models, it is natural to ask whether these predictions can reduce the number of costly labels needed to answer scientific questions of interest, e.g., what is the prevalence of this disease or the change in cell-type proportions across different experimental conditions. 

Recent work has shown that treating predicted labels as direct observations is invalid and inflates Type~I error. Methods such as \emph{Prediction-Powered Inference}
(PPI)~\citep{angelopoulos2023prediction} and its tuned variant
\texttt{PPI++}~\citep{angelopoulos2023ppipp} instead calibrate predictions against a small set of gold-standard labels. Together with related work~\citep{zrnic2024cross,miao2025assumption,egami2023using,gronsbell2024another}, these methods yield unbiased and asymptotically more efficient estimators than classical analyses based on gold-standard labels alone. However, less work has translated those efficiency gains into the design-stage question every study planner must answer:
\begin{tcolorbox}
\emph{Given a predictive model's accuracy, how many labeled samples
are needed to detect an effect of interest with the desired
statistical power?}
\end{tcolorbox}

Classical power formulas~\citep{erdfelder1996gpower,faul2007gpower} do not provide a straightforward way to distinguish between gold-standard labels and predictions. As a result, practitioners either ignore predictions entirely, which leads to valid planning but potential losses in efficiency, or naively treat predictions as gold-standard labels, which leads to overestimated power and underestimated sample sizes. Several recent contributions start to address this gap: \citet{angelopoulos2023ppipp} develop cost-allocation routines from pilot data but do not provide closed-form power or sample-size formulas.  Comprehensive software packages~\citep{salerno2025ipd,egami2024using} unify a range of post-prediction inference methods yet focus on the analysis stage
and offer no prospective study-design functionality. \citet{poulet2025prediction} derive power reductions for
ANCOVA-style regression adjustment but restrict attention to that single estimator. While most work focused on comparing asymptotic efficiency of estimators, \citet{mani2025no} also investigated the finite-sample trade-off, and our rule of thumb gives that limitation a concrete planning interpretation. Closest to our work is \citet{broska2025mixed}, who invert the \texttt{PPI++} variance formula to jointly optimize the labeled and unlabeled sample sizes under a cost constraint.  Our starting point differs in a practical but important way: in many biomedical settings the two sample sizes are \emph{not} jointly optimized.  The pool of unlabeled observations is often fixed by the experiment itself (e.g., the number of cells captured in a single-cell assay, or the patients meeting inclusion criteria), and the binding constraint is how many a biologist or physician can manually label.  We therefore condition on a given~$N$ and ask how many gold-standard labels~$n$ are needed to achieve target power.

In this paper, we propose \texttt{pppower}, a framework and R package for prediction-powered power analysis (see Figure~\ref{fig:intro-nhanes}).  Our contributions are as follows: 
\begin{enumerate}
  \item \textbf{Closed-form formulas.}\  We derive power and
    sample-size formulas for prediction-powered estimators that
    combine a smaller set of gold-standard labels with a larger pool
    of model-based predicted labels, and show that they reduce to the
    classical formulas when the predictions are uninformative.  These
    formulas cover two-sample comparisons, paired
    designs, odds ratios and relative risks in $2\times 2$ tables,
    and regression contrasts for (generalized) linear models
    (Propositions~\ref{prop:ppi-var}--\ref{prop:paired-var},
    \ref{prop:2x2-n}, and~\ref{prop:regression-n}).
    A simple rule of thumb emerges: the required labeled sample size
    drops by approximately $R^{2}\times 100\%$.
  \item \textbf{User-friendly parameterization.}\  The formulas
    accept $R^{2}$, sensitivity/specificity, or confusion matrices
    as inputs for the one-sample and binary-table designs, avoiding
    direct covariance specification in the main operational cases;
    regression-contrast extensions still require contrast-level pilot
    covariance blocks (Sections~\ref{sec:calibration}
    and~\ref{sec:regression-main}).
  \item \textbf{Simulation validation.}\  A comprehensive
    simulation study validates all formulas against Monte~Carlo
    estimates and establishes practical guidance on when model-based
    predicted labels yield the largest reductions in required labeled
    sample size (Section~\ref{sec:simulation}).
\item \textbf{Open-source software with case study.}\  An R package with a
    \texttt{pwr}-style API and an accompanying online calculator unify power
    calculation and sample-size determination in a single
    function call. We demonstrate the package on three biomedical case studies
    (Section~\ref{sec:applications}).
\end{enumerate}

\PPIIntroFigurePlacement

\section{Setup and Background}
\label{sec:setup}

We consider i.i.d.\ labeled data $\{(X_i, Y_i)\}_{i=1}^{n}$ drawn from a
joint distribution $P_{XY}$ on $\mathcal{X} \times \mathcal{Y}$, and an
independent pool of unlabeled covariates $\{\wt{X}_j\}_{j=1}^{N}$ drawn from
the same marginal $P_X$, with $r = n/N$ denoting the labeled-to-unlabeled
ratio. Throughout, $n$ denotes the number of gold-standard labels and $N$ the size of the larger unlabeled sample carrying model-based predictions, with the practically important regime often having $N \gg n$. 
A predictor $f : \mathcal{X} \to \mathcal{Y}$ yields predictions
$f_i = f(X_i)$ on labeled observations and $\wt{f}_j = f(\wt{X}_j)$ on
unlabeled ones; we treat $f$ as independent of both samples (e.g., a
pre-trained model), and note that cross-fitting can relax this assumption in
practice (see the Supplementary Material and \citet{zrnic2024cross}).

Beyond $\sigma_Y^2 = \Var(Y)$, three second-order quantities govern the
interplay between outcomes and predictions throughout the paper. The prediction
variance $\sigma_f^2 = \Var(f(X))$ and the residual variance
$\sigma_\varepsilon^2 = \Var(Y - f(X)) = \sigma_Y^2 + \sigma_f^2 - 2\Cov(Y,f)$
together measure how much variation in $Y$ the predictor captures; the
outcome--prediction correlation $\rho_{Yf} = \Cov(Y,f)/(\sigma_Y\sigma_f)$
summarizes this in a single unitless quantity, with $\rho_{Yf}^2$ equal to the
fraction of $\sigma_Y^2$ explained by $f$. Finally, we define the variance
threshold
\begin{equation}\label{eq:s2-threshold}
  S^2 \;=\; \Bigl\{\Delta \,/\, (z_{1-\alpha/2}+z_{1-\beta})\Bigr\}^2.
\end{equation}
\subsection{Testing for One-Sample Mean}
\label{sec:classical}

We first consider testing equality of the population mean $\theta^\star = \E[Y]$:
\begin{equation}\label{eq:test}
  H_0 : \theta^\star = \theta_0
  \quad\text{versus}\quad
  H_1 : \theta^\star = \theta_0 + \Delta,
\end{equation}
at significance level $\alpha$ with target power $1-\beta$. The sample mean
$\bar{Y} = n^{-1}\sum_{i=1}^n Y_i$ is unbiased for $\theta^\star$ with
variance $\sigma_Y^2/n$. The power of the two-sided Wald test
for~\eqref{eq:test} is
\begin{equation}\label{eq:classical-power}
  \mathrm{Power}_{\mathrm{cl}}(n)
  = \Phi\!\left(-z_{1-\alpha/2} + \frac{|\Delta|}{\sigma_Y/\sqrt{n}}\right)
  + \Phi\!\left(-z_{1-\alpha/2} - \frac{|\Delta|}{\sigma_Y/\sqrt{n}}\right).
\end{equation}
Inverting~\eqref{eq:classical-power} by dropping the negligible second term gives $n_{\mathrm{cl}} \ge \sigma_Y^2 / S^2$. For fixed $\alpha$, $\beta$, and $\Delta$, $\sigma_Y^2$ is the only determinant of the required sample
size: reducing it is the only way to increase power for a fixed $n$.
\subsection{Recap of Prediction-Powered Inference}
\label{subsec:ppi-power}\label{sec:ppi-vanilla}
The classical power formula in \eqref{eq:classical-power} cannot directly make use of the predictions $\tilde{f}$.  In this section, we first review the PPI and \texttt{PPI++} estimators and their variances, which can then be inverted to obtain the power formulas using both gold-standard labels and predictions.

When the population mean is the target of interest, the PPI estimator~\citep{angelopoulos2023prediction} makes a simple observation that the population mean decomposes as $\theta^\star = \E[f(X)] + \E[Y - f(X)]$, and each component can be estimated separately:
\begin{equation}\label{eq:ppi-estimator}
  \wh{\theta}_{\mathrm{PPI}}
  = \underbrace{\frac{1}{N}\sum_{j=1}^N \wt{f}_j}_{\text{unlabeled mean}}
  \;+\;
  \underbrace{\frac{1}{n}\sum_{i=1}^n (Y_i - f_i)}_{\text{labeled correction}}.
\end{equation}
\begin{proposition}[Mean \texttt{PPI++} variance]\label{prop:ppi-var}\label{prop:ppi++-var}
Under our setup, the PPI estimator is unbiased and asymptotically normal with variance
\begin{equation}\label{eq:ppi-var}
  \Var(\wh{\theta}_{\mathrm{PPI}})
  = \frac{\sigma_f^2}{N} + \frac{\sigma_\varepsilon^2}{n}.
\end{equation}
The two terms reflect prediction noise $\sigma_f^2/N$ from the unlabeled sample and residual
noise $\sigma_\varepsilon^2/n$ from the labeled sample. When $N$ is large, the variance is dominated by $\sigma_\varepsilon^2/n$.

While PPI in \eqref{eq:ppi-estimator} uses $f_i$, the resulting estimator
remains unbiased if we replace $f_i$ with any function $g(f_i)$, and an
appropriate choice of $g$ could further reduce the variance.
\texttt{PPI++}~\citep{angelopoulos2023ppipp} considers $g(f_i) = \lambda f_i$
with a tunable parameter $\lambda \in \mathbb{R}$ controlling the weight placed
on predictions:
\begin{equation}\label{eq:ppi++-estimator}
  \wh{\theta}_\lambda = n^{-1}\sum_{i=1}^n Y_i + \lambda\,(N^{-1}\sum_{j=1}^N
  \wt{f}_j - n^{-1}\sum_{i=1}^n f_i).
\end{equation}
The asymptotic variance of $\wh{\theta}_\lambda$ is
\begin{equation}\label{eq:ppi++-variance}
  \Var(\wh{\theta}_\lambda)
  = \frac{\sigma_Y^2}{n}
  + \lambda^2\sigma_f^2\!\left(\frac{1}{N}+\frac{1}{n}\right)
  - \frac{2\lambda\,\Cov(Y,f)}{n},
\end{equation}
and the asymptotic-variance-minimizing tuning parameter is
\begin{equation}\label{eq:lambda-star}
  \lambda^\star = \frac{\Cov(Y,f)}{(1+r)\,\sigma_f^2},
\end{equation}
with resulting optimal variance
\begin{equation}\label{eq:ppi++-var-opt}
  \Var(\wh{\theta}_{\lambda^\star})
  = \frac{\sigma_Y^2}{n}
  - \frac{\Cov(Y,f)^2}{\sigma_f^2}\cdot\frac{N}{n(n+N)}.
\end{equation}
\end{proposition}

\section{Power and Sample Size for \texttt{PPI++}}
\label{sec:ppi-power-formula}\label{sec:ppi-power}\label{sec:ppi++}

Since PPI is the special case of \texttt{PPI++} with $\lambda = 1$, we state
all results for general $\lambda$; setting $\lambda = 1$ recovers the
PPI-specific results. The optimal variance~\eqref{eq:ppi++-var-opt} plays the
same role for \texttt{PPI++} that $\sigma_Y^2/n$ plays for the classical
estimator: it determines the power of the Wald test.

\subsection{Tests for a One-Sample Mean}

\begin{proposition}[\texttt{PPI++} power]\label{prop:ppi++-power}
Consider testing~\eqref{eq:test} using the Wald statistic
$Z = (\wh{\theta}_{\lambda^\star} - \theta_0)
/ \sqrt{\Var(\wh{\theta}_{\lambda^\star})}$.
Recalling $\Var(\wh{\theta}_{\lambda^\star})$ from~\eqref{eq:ppi++-var-opt}, the power of the two-sided test at level $\alpha$ is
\begin{equation}\label{eq:ppi++-power}
\mathrm{Power}_{\mathrm{PPI}}(n, N)
= \Phi\!\left(-z_{1-\alpha/2}
  + \frac{|\Delta|}{\sqrt{\Var(\wh{\theta}_{\lambda^\star})}}\right)
+ \Phi\!\left(-z_{1-\alpha/2}
  - \frac{|\Delta|}{\sqrt{\Var(\wh{\theta}_{\lambda^\star})}}\right).
\end{equation}
\end{proposition}

Inverting~\eqref{eq:ppi++-power} by setting
$\Var(\wh{\theta}_{\lambda^\star}) \le S^2$ and solving the resulting quadratic
in $n$ yields the following sample-size formula.

\begin{proposition}[\texttt{PPI++} sample size]\label{prop:ppi++-n}
The minimum labeled sample size required to achieve power $1-\beta$ is
\begin{equation}\label{eq:ppi++-n-formula}
n \;\ge\;
\frac{\sigma_Y^2 - S^2 N
  + \sqrt{(\sigma_Y^2 - S^2 N)^2
    + 4 S^2 N \sigma_Y^2 (1 - \rho_{Yf}^2)}}
{2 S^2}.
\end{equation}
\end{proposition}

We write $n^\star$ for the smallest integer labeled sample size
satisfying~\eqref{eq:ppi++-n-formula}, that is, the inverted minimum
number of gold-standard labels needed to attain the target power. For
fixed $N$, the inversion in~\eqref{eq:ppi++-n-formula} can return
$n^\star > N$. This does not invalidate the variance formula; it simply means that a fixed-pool design would exhaust the available
prediction pool. Such a regime may be of limited practical interest,
as the motivating assumption is that the main constraint lies in the
cost of obtaining gold-standard labels.

\begin{corollary}[Rule of thumb]\label{cor:rule-of-thumb}
In the regime $N \gg n$, the term $\sigma_f^2/N$ becomes negligible and
the optimal variance reduces to
$\Var(\wh{\theta}_{\lambda^\star}) \approx \sigma_Y^2(1 - \rho_{Yf}^2)/n$,
yielding the simple approximation
$n_{\mathrm{PPI}} / n_{\mathrm{cl}} \approx 1 - \rho_{Yf}^2$.
\end{corollary}

\begin{tcolorbox}
\textbf{Rule of thumb.}~A predictor with $R^2 = \rho_{Yf}^2$ reduces the
required labeled sample size by approximately $R^2 \times 100\%$. For
example, $R^2 = 0.5$ halves the labeled-data requirement, and $R^2 = 0.9$
yields up to a $90\%$ reduction.
\end{tcolorbox}

\subsection{Connection to Semiparametric Efficiency}
\label{sec:eif}

The family of estimators
$n^{-1}\sum_{i=1}^n Y_i - n^{-1}\sum_{i=1}^n g(f_i) + N^{-1}\sum_{j=1}^N g(\wt{f}_j)$
is unbiased for $\theta^\star$ for any function $g$, and the choice of $g$
governs the asymptotic variance. For the mean functional, choosing
$g(f) = \E[Y \mid f]$ minimizes the asymptotic variance not merely within
this family but among all asymptotically unbiased estimators of
$\theta^\star$, a direct consequence of classical semiparametric efficiency
theory. In this one-sample mean setting, the factor $1 - \rho_{Yf}^2$ in
the rule of thumb above represents the fundamental limit on how much
predictions can help: it is the efficiency bound, not just the performance of
a particular estimator.

In the special case where $Y$ and $f(X)$ are both binary, the linear
construction with optimal $\lambda$ achieves this bound, so
\texttt{PPI++} is semiparametrically efficient. In more general settings the
linear form may be suboptimal~\citep{chen2026unifying, ji2025predictions,xu2025unified,song2026demystifying}, and the variance-optimal estimator could replace $\lambda f_i$
with a non-linear estimate $\wh{g}(f_i)$ of $\E[Y \mid f]$, which can be
estimated from the labeled data. In this case, the same set of arguments holds but the estimated variance will differ.

\subsection{Calibration from Common Metrics}
\label{sec:calibration}

For the one-sample mean formulas above, the prediction enters through
$\Cov(Y,f)^2/\sigma_f^2 = \sigma_Y^2\rho_{Yf}^2$, so the key planning inputs
reduce to the outcome variance $\sigma_Y^2$ and the squared
outcome--prediction correlation $\rho_{Yf}^2$. In practice, however, ML model
documentation does not always report these quantities directly: continuous
models are typically summarized by held-out $R^2$, while classifiers are
characterized by metrics such as accuracy, precision, and recall. 

\paragraph{Continuous outcomes.}
When the model $R^2$ is reported in prior publications, we can use $R^2 = \rho_{Yf}^2$ directly. Many
software packages instead report the mean-squared error
$\sigma_\varepsilon^2 = \sigma_Y^2 + \sigma_f^2 - 2\Cov(Y,f)$. This summary
does not identify $\sigma_f^2$ and $\Cov(Y,f)$ separately, but it still gives
a conservative planning input: by Cauchy--Schwarz,
$\sigma_\varepsilon^2 \geq \sigma_Y^2(1-\rho_{Yf}^2)$, so
$R^2 \leq \rho_{Yf}^2$. Thus plugging such an $R^2$ into the rule of thumb,
or into \eqref{eq:ppi++-n-formula} through $\rho_{Yf}^2 \approx R^2$, will give us a more conservative estimate of the number of required samples.

\paragraph{Binary outcomes.}
\label{sec:sens-spec}
For a binary outcome $Y \in \{0,1\}$ and a binary classifier $f \in
\{0,1\}$, the required variance components are determined by three quantities:
the outcome prevalence $p= P(Y=1)$, the sensitivity $\mathrm{se} = P(f=1
\mid Y=1)$ and the specificity $\mathrm{sp} = P(f=0 \mid Y=0)$. From these, the prediction
prevalence is $p_f = \mathrm{se}\cdot p + (1-\mathrm{sp})(1-p)$, and the
three inputs follow as $\sigma_Y^2 = p(1-p)$, $\sigma_f^2 = p_f(1-p_f)$, and
$\Cov(Y,f) = \mathrm{se}\cdot p - p\cdot p_f$.

\section{Extensions to Other Designs}
\label{sec:extensions}

We next extend the one-sample framework of Section~\ref{sec:ppi-power-formula} to
other commonly used designs: two-sample comparisons (unpaired and paired), $2\times2$ table tests, and contrasts for linear and logistic models.

\subsection{Tests for Equality of Two Means}
\label{sec:ttest}

Consider comparing two independent groups $A$ and $B$ with population means
$\theta_A^\star$ and $\theta_B^\star$, where the parameter of interest is
$\theta_A^\star - \theta_B^\star$. For each group $g \in \{A, B\}$, we observe
$n_g$ labeled pairs and $N_g$ unlabeled observations, and construct a
group-specific \texttt{PPI++} estimator $\wh{\theta}_{g,\mathrm{PPI}}$ as
in~\eqref{eq:ppi++-estimator}. The \texttt{PPI++} estimator of the group
difference is
\begin{equation}\label{eq:ppi++-ttest-est}
  \wh{\Delta}_{\mathrm{PPI}} = \wh{\theta}_{A,\mathrm{PPI}} -
  \wh{\theta}_{B,\mathrm{PPI}}.
\end{equation}

\begin{proposition}[Two-sample \texttt{PPI++} variance]\label{prop:ttest-var}
Under independent groups, the variance of $\wh{\Delta}_{\mathrm{PPI}}$ at the
group-specific oracle tuning parameters is
\begin{equation}\label{eq:ppi++-ttest-var}
\Var(\wh{\Delta}_{\mathrm{PPI}})
= \sum_{g \in \{A,B\}}
  \left[\frac{\sigma_{Y,g}^2}{n_g}
  - \frac{\Cov(Y_g, f_g)^2}{\sigma_{f,g}^2}
    \cdot \frac{N_g}{n_g(n_g + N_g)}\right].
\end{equation}
In the regime $N_A, N_B \gg n_A, n_B$, this reduces to
\begin{equation}\label{eq:ppi++-ttest-var-approx}
\Var(\wh{\Delta}_{\mathrm{PPI}})
\;\approx\;
\frac{\sigma_{Y,A}^2\bigl(1 - \Corr(Y_A,f_A)^2\bigr)}{n_A}
+\frac{\sigma_{Y,B}^2\bigl(1 - \Corr(Y_B,f_B)^2\bigr)}{n_B},
\end{equation}
which is the sum of the group-specific one-sample variances, each reduced by
the squared outcome--prediction correlation within that group.
\end{proposition}

The required sample sizes $n_A$ and $n_B$ follow from setting
$\Var(\wh{\Delta}_{\mathrm{PPI}}) \le S^2$ from~\eqref{eq:s2-threshold} and specifying an allocation ratio
$n_A/n_B$; for a balanced design ($n_A = n_B = n$) the condition becomes $n
\ge [\sigma_{Y,A}^2(1-\Corr(Y_A,f_A)^2) +
\sigma_{Y,B}^2(1-\Corr(Y_B,f_B)^2)] / S^2$.

\subsection{Tests for Paired Mean Differences}
\label{sec:paired}

In paired designs, each subject $i$ contributes measurements under both
conditions (e.g., before and after treatment), yielding within-subject
differences $Y_i^A - Y_i^B$ with paired predictions $f^A(X_i) - f^B(X_i)$.
The parameter of interest is $\E[Y^A - Y^B]$, and the problem reduces
directly to the one-sample framework applied to the differences.

\begin{proposition}[Paired \texttt{PPI++} variance]\label{prop:paired-var}
Let $D_i = Y_i^A - Y_i^B$ and $G_i = f_i^A - f_i^B$. Applying the
one-sample \texttt{PPI++} result to the paired differences gives
\[
\Var(\wh{\Delta}_{\mathrm{PPI++}})
=
\frac{\Var(D)}{n}
-
\frac{\Cov(D,G)^2}{\Var(G)}
\cdot
\frac{N}{n(n+N)}
=
\frac{\Var(D)}{n}
\left(
1-\Corr(D,G)^2\frac{N}{n+N}
\right).
\]
When $N \gg n$, this becomes
$\Var(\wh{\Delta}_{\mathrm{PPI++}})
\approx \Var(Y^A-Y^B)\{1-\Corr(Y^A-Y^B,\;f^A-f^B)^2\}/n$.
\end{proposition}

Just as a paired design can be more efficient than an unpaired two-sample test
by removing between-subject noise, the correlation $\Corr(Y^A - Y^B, f^A -
f^B)$ governs the efficiency gain here and can differ substantially from the
marginal correlations $\Corr(Y_A, f_A)$ and $\Corr(Y_B, f_B)$: if the
predictor captures within-subject variation well, this correlation can be high
even when the marginal correlations are moderate, and conversely.

\subsection{Tests for Relative Risk and Odds Ratio in $2\times2$ Tables}
\label{sec:2x2}

Many clinical and epidemiological studies compare whether a binary event occurs
in two groups, such as treatment ($g=1$) and control ($g=0$). For each group
$g$, the labeled data contain event indicators $Y_{gi} \in \{0,1\}$, while the
prediction model produces a surrogate $f_g(X)$ for that same event indicator on
both the labeled and unlabeled subjects in that group. This setup is often summarized as a $2\times2$ table of aggregated counts indexed by the group label $g$ and the outcome $Y$.

The two standard effect measures in this setting are
$\mathrm{RR} = \frac{p_1}{p_0}, \mathrm{OR} = \frac{p_1/(1-p_1)}{p_0/(1-p_0)}$,
and under the null hypothesis that the treatment $g$ has no effect, we would have $\mathrm{RR}=\mathrm{OR} = 1$. In practice, we often consider the equivalent formulation on the log scale using
$\log(\mathrm{RR}) = \log p_1 - \log p_0$ and
$\log(\mathrm{OR}) = \operatorname{logit}(p_1) - \operatorname{logit}(p_0)$. 

As both estimands only depend on the group-specific event probability
$p_g = P(Y=1 \mid G=g)$, it suffices to estimate $p_g$ using
$\wh{p}_{g,\lambda_g} = \bar{Y}_{L,g} +
\lambda_g(\bar{f}_{U,g} - \bar{f}_{L,g})$.
Here, $\bar{Y}_{L,g}$ is the labeled event rate, and $\bar{f}_{L,g}$ and
$\bar{f}_{U,g}$ are the average predictions in the labeled and unlabeled
samples from group $g$.  Under our independence assumption, the
asymptotic variance of $\log \widehat{\mathrm{RR}}$ and
$\log \widehat{\mathrm{OR}}$ follows directly from the delta method.

\begin{proposition}[\texttt{PPI++} sample size for relative risk and odds ratio]\label{prop:2x2-n}
Assume the two groups are independent, let $\rho_g = \Corr(Y_g, f_g)$,
and fix an allocation ratio $\kappa = n_1/n_0$. In the regime
$N_g \gg n_g$, define
$\Delta_{\mathrm{RR}} = \log(p_1/p_0)$,
$\Delta_{\mathrm{OR}} = \operatorname{logit}(p_1) -
\operatorname{logit}(p_0)$,
$S_{\mathrm{RR}}^2 =
\Delta_{\mathrm{RR}}^2/(z_{1-\alpha/2}+z_{1-\beta})^2$, and
$S_{\mathrm{OR}}^2 =
\Delta_{\mathrm{OR}}^2/(z_{1-\alpha/2}+z_{1-\beta})^2$. Then
\begin{align*}
\Var(\log \widehat{\mathrm{RR}})
&\approx
\frac{1}{n_0}
\left[
\frac{(1-p_0)(1-\rho_0^2)}{p_0}
+
\frac{1}{\kappa}\frac{(1-p_1)(1-\rho_1^2)}{p_1}
\right], \\
\Var(\log \widehat{\mathrm{OR}})
&\approx
\frac{1}{n_0}
\left[
\frac{1-\rho_0^2}{p_0(1-p_0)}
+
\frac{1}{\kappa}\frac{1-\rho_1^2}{p_1(1-p_1)}
\right].
\end{align*}
Consequently, the minimum labeled sample sizes required for a two-sided Wald
test at level $\alpha$ with power $1-\beta$ are
\begin{align*}
n_0^{\mathrm{RR}}
&\ge
\frac{1}{S_{\mathrm{RR}}^2}
\left[
\frac{(1-p_0)(1-\rho_0^2)}{p_0}
+
\frac{1}{\kappa}\frac{(1-p_1)(1-\rho_1^2)}{p_1}
\right],
&
n_1^{\mathrm{RR}} &= \kappa n_0^{\mathrm{RR}}, \\
n_0^{\mathrm{OR}}
&\ge
\frac{1}{S_{\mathrm{OR}}^2}
\left[
\frac{1-\rho_0^2}{p_0(1-p_0)}
+
\frac{1}{\kappa}\frac{1-\rho_1^2}{p_1(1-p_1)}
\right],
&
n_1^{\mathrm{OR}} &= \kappa n_0^{\mathrm{OR}}.
\end{align*}
For balanced designs, set $\kappa = 1$.  Setting $\rho_0 = \rho_1 = 0$
recovers the classical relative-risk and odds-ratio sample-size
formulas.
\end{proposition}
\subsection{Tests for Regression Contrasts in Generalized Linear Models}
\label{sec:regression-main}
Many of the power analyses above can be recast as testing a linear
contrast in a generalized linear model. We first motivate
prediction-powered inference for linear models. Consider the ordinary
least squares (OLS) estimator with target coefficient vector
$\beta^\star$ that minimizes $\E[(Y - X^\top\beta)^2]$. The normal
equations imply $\E[XX^\top]\beta^\star = \E[XY]$. 

Now, for covariate--outcome observation pairs $(x_i,y_i)$, if $f(x_i)$ is a
prediction of $y_i$, we can decompose the normal equation as
$\E[XY] = \E[Xf(X)] + \E[X\{Y-f(X)\}],$
which leads to the estimating equation
\[
\begin{aligned}
\frac{1}{n}\sum_{i=1}^n X_i\big(Y_i - X_i^\top\beta\big) +
\lambda\left[
\frac{1}{N}\sum_{j=1}^N \wt{X}_j\big(\wt{f}_j - \wt{X}_j^\top\beta\big)
-
\frac{1}{n}\sum_{i=1}^n X_i\big(f_i - X_i^\top\beta\big)
\right] = 0.
\end{aligned}
\]

Similarly, for a generalized linear model with mean function $\mu_\beta(X)$,
the target parameter satisfies the score equation
$\E[X\{Y-\mu_\beta(X)\}] = 0$ at $\beta=\beta^\star$, motivating the decomposition
\[
\E[X\{Y-\mu_\beta(X)\}]
=
\E[X\{\mu_f(X)-\mu_\beta(X)\}]
\;+\;
\E[X\{Y-\mu_f(X)\}],
\]
which yields the analogous \texttt{PPI++} score equation
\[
\begin{aligned}
\frac{1}{n}\sum_{i=1}^n X_i\big(Y_i-\mu_\beta(X_i)\big) +
\lambda\left[
\frac{1}{N}\sum_{j=1}^N \wt{X}_j\big(\mu_f(\wt{X}_j)-\mu_\beta(\wt{X}_j)\big)
-
\frac{1}{n}\sum_{i=1}^n X_i\big(\mu_f(X_i)-\mu_\beta(X_i)\big)
\right]=0.
\end{aligned}
\]

With $r=n/N$, a first-order expansion of $\wh{\theta}_\lambda$
gives
\[
\Var(\wh{\theta}_\lambda)
\;\approx\;
\frac{V_{YY} + \lambda^2(1+r)V_{ff} - 2\lambda V_{Yf}}{n},
\]
where $V_{YY}$, $V_{ff}$, and $V_{Yf}$ denote the contrast-level
variances of the labeled score term, the prediction term, and their
covariance.

Below, in Proposition~\ref{prop:regression-n}, we summarize the power and required sample
size for the contrast of interest $\theta^\star = a^\top \beta^\star$
using the \texttt{PPI++} estimator.

\begin{proposition}[\texttt{PPI++} sample size for regression contrasts]\label{prop:regression-n}
Let $\Delta = a^\top\beta^\star$ denote the target contrast under the
alternative. The oracle tuning parameter is
$\lambda^\star = V_{Yf}/\{(1+r)V_{ff}\}$, which yields
\[
\Var(\wh{\theta}_{\lambda^\star})
\;\approx\;
\frac{1}{n}\left(V_{YY} - \frac{V_{Yf}^2}{(1+r)V_{ff}}\right).
\]
Treating $N$ as fixed, the minimum labeled sample size required for a two-sided
Wald test at level $\alpha$ with power $1-\beta$ satisfies
\[
n \ge
\frac{
V_{YY} - S^2N
+
\sqrt{(V_{YY}-S^2N)^2 + 4S^2N\left(V_{YY} - \frac{V_{Yf}^2}{V_{ff}}\right)}
}{2S^2}.
\]
In the regime $N \gg n$, this simplifies to $
n \ge \frac{V_{YY} - V_{Yf}^2/V_{ff}}{S^2}$.
\end{proposition}

\section{Simulation Studies}
\label{sec:simulation}

Here, we validate the approximate closed-form formulas derived in Sections~\ref{sec:ppi-power}--\ref{sec:extensions} through comprehensive Monte Carlo simulations. For each scenario, we generate $R = 1{,}000$ replicates and compare the empirical power of the \texttt{PPI++}-based tests with the theoretical formulas we presented. Unless otherwise noted, all tests are two-sided at level $\alpha = 0.05$ and use the oracle $\lambda^\star$, and we target a power of $80\%$ in the sample size calculations. We also examine several robustness checks, including whether the \texttt{PPI++}-based tests have the correct size (i.e., control Type I error at the nominal level), the discrepancy between the estimated and oracle values of $\lambda^\star$, and departures from the normality assumptions (e.g., small $n$ or non-normal outcomes). We present the main power and sample size validation results in the main text, with additional robustness checks deferred to the Supplementary Material.

\subsection{Power Validation}
\label{sec:sim-core}

\paragraph{Mean estimation.}
We consider both continuous and binary outcomes. For the continuous case, we simulate $(Y, f)$ as bivariate normal with $\sigma_Y^2 = \sigma_f^2 = 1$, correlation $\rho \in \{0.5, 0.7, 0.9\}$, and
$(n, N) \in \{20, 40, 60, 80, 100\} \times \{200, 500\}$, with a target effect size of $\Delta = 0.2$. For the binary outcome experiment, we generate outcomes $Y \sim \mathrm{Bernoulli}(p = 0.3)$ with predictions generated via sensitivity and specificity conditional on $Y$, and $\Delta = 0.05$.

Across both settings, power increases as the predictive accuracy of the labels increases (higher $\rho$) and as the number of labeled samples ($n$) increases, and there is strong agreement between the theoretical and empirical power (Figure~\ref{fig:panels-AB}). For binary outcomes, the analytical approximation captures the overall trend, but the smallest-$n$, highest-accuracy settings show visible finite-sample departures (e.g., sensitivity $=$ specificity $= 0.95$, $N = 500$, $n = 20$). Our power planning formula is mildly optimistic (about 10\% inflation), primarily due to instability in the estimated denominator at very small $n$.

\PPIMeanValidationFigurePlacement

\paragraph{Two-sample tests.}
We simulate independent groups with continuous ($\Delta = 0.3$) and
binary ($\Delta = 0.08$) outcomes using the same parameter grids as in
the mean estimation experiments. Across both outcome types, the
theoretical curves track the empirical power closely over the full
$(n, N)$ grid, with maximum absolute discrepancies of 0.02 and 0.03 for
continuous and binary outcomes, respectively. The largest deviations
typically occur in the most prediction-favorable regimes with very
small labeled samples, where the denominator in the estimated variance
of the Wald test is again the main source of extra variability. See
Figures~\ref{fig:setting-C}--\ref{fig:setting-D} in the Supplementary
Material.

\paragraph{Paired designs.}
Paired continuous and binary designs with within-pair prediction
correlation show the same overall pattern of strong agreement between
theoretical and empirical power (see Figures~\ref{fig:setting-E}--\ref{fig:setting-F}), with the largest gaps again appearing at small $n$. These results suggest that the paired-design approximation remains accurate once the within-pair covariance structure is accounted for.

\subsection{Extensions: $2\times2$ Tables and Generalized Linear Models}
\label{sec:sim-regression}

We next validate the contingency-table and regression extensions
developed in Section~\ref{sec:extensions}.

\paragraph{$2\times 2$ table (odds ratio and relative risk).}
The data-generating process draws binary treatment $X \in \{0,1\}$ and
binary outcome $Y$ with control probability $p_{\mathrm{ctrl}} = 0.20$ and experimental probability
$p_{\mathrm{exp}} \in \{0.30, 0.35, 0.40\}$.  A noisy binary
classifier with sensitivity $=$ specificity $=$ accuracy serves as
the surrogate, with accuracy $\in \{70\%, 80\%, 90\%\}$. Figure~\ref{fig:setting-U} 
displays the power result for $p_{\mathrm{exp}} = 0.40$, which corresponds to a relative risk of $2.00$ and odds ratio of $2.67$; agreement is generally tight.

\PPIBinaryTableFigurePlacement

\paragraph{Linear regression contrast.}
We simulate $X \sim N(0, I_2)$, $Y = X\beta + \varepsilon$ with
$\beta = (\Delta, 0)$, $\Delta = 0.3$, and predictions
$f = X\beta + \nu$ where
$\nu = \rho\,\varepsilon + \sqrt{1 - \rho^2}\,\eta$,
$\eta \sim N(0,1)$.  The contrast $a = (1, -1)$ tests
$H_0\!: \beta_1 - \beta_2 = 0$.  Under this simulation setting, $\Var(a^\top\beta)$ in Proposition~\ref{prop:regression-n} reduces to $V_{YY} = 2$, $V_{ff} = 2$, and $V_{Yf} = 2\rho$.
Each Monte Carlo replicate constructs the \texttt{PPI++} estimator with
oracle $\lambda^\star$ and a sandwich estimator for the standard error, then performs a
two-sided Wald test.  Figure~\ref{fig:setting-ST}(a) shows close
agreement across all 30 configurations
($n \in \{20, 40, 60, 80, 100\}$, $N \in \{200, 500\}$,
$\rho \in \{0.5, 0.7, 0.9\}$); the maximum discrepancy is~0.08 at small~$n$ where finite-sample sandwich bias
is expected.

\paragraph{Logistic regression contrast.}
We extend to non-linear models by simulating $X \sim N(0, I_2)$,
$P(Y {=} 1 \mid X) = \operatorname{expit}(X\beta)$ with
$\beta = (0.5, 0)$ (corresponding to an odds ratio of $e^{0.5} \approx 1.65$).
We consider a noisy binary classifier with sensitivity $=$ specificity $=$
accuracy as the surrogate. Because the required contrast-level variance
blocks do not admit convenient closed forms for logistic regression, we
approximate the corresponding population quantities once using a large
Monte Carlo reference sample ($M = 100{,}000$) and use them to evaluate
the analytical power curve. For the empirical power calculation, each
fresh simulated dataset estimates a two-fold cross-fitted plug-in
tuning parameter $\hat\lambda$ and then solves the corresponding
rectified score equation using that plug-in estimate.
Figure~\ref{fig:setting-ST}(b) displays results across 30 configurations
($n \in \{20, 40, 60, 80, 100\}$, $N \in \{200, 500\}$,
accuracy $\in \{70\%, 80\%, 90\%\}$); the maximum discrepancy
is~0.1, concentrated at small~$n$ with high accuracy where estimating
$\hat\lambda$ adds the most finite-sample variability.

\PPIRegressionFigurePlacement

\subsection{Robustness Checks}
\label{sec:sim-robustness}

We also examine sample-size inversion accuracy, Type I error control of the \texttt{PPI++} tests, and departures from Gaussian assumptions. Fixing target
power in $\{0.60, 0.70, 0.80, 0.90\}$ shows that the planned integer
sample size $n^\star$ delivers achieved power close to target across
the one-sample, two-sample, and paired designs; analytical deviations
are uniformly small, and empirical deviations are only modest in the
smallest-$n$ cells (Figure~\ref{fig:setting-J} in the Supplementary
Material). We also verified that, under the null, rejection rates remain close to the nominal
$0.05$ level, ranging from $0.04$ to $0.06$ in the continuous
and binary settings (Figure~\ref{fig:setting-L} in the Supplementary Material). Moreover, with
$N = 1{,}000$ fixed, the plug-in estimate of the tuning parameter $\hat\lambda$ converges
steadily toward the oracle $\lambda^\star$ as $n$ grows
(Figure~\ref{fig:setting-M} in the Supplementary Material).

We next stress the formulas through practical implementation choices
and finite-sample perturbations around the Gaussian benchmark.
Replacing the oracle tuning parameter by the plugin estimate
$\hat\lambda$ changes power very little on the Gaussian one-sample settings: the maximum oracle--plugin difference is $0.03$, and both remain close to the analytical curve
(Figure~\ref{fig:setting-N} in the Supplementary Material). A sweep
over effect size $\Delta \in [0, 0.5]$ at fixed $N = 500$ produces the
expected S-shaped power curves, with larger $\rho$ shifting the
\texttt{PPI++} curves leftward (Figure~\ref{fig:setting-O} in the
Supplementary Material). Pushing the Gaussian one-sample design down to
$n \in \{15, 20, 25, 30, 50, 100\}$ shows that the approximation
remains usable but is predictably weakest below about $n = 25$
(Figure~\ref{fig:setting-P} in the Supplementary Material), while
varying $N/n$ from $1$ to $100$ at fixed $n = 50$ shows that gains from
unlabeled data rise quickly and then saturate around
$N/n \approx 10$--$20$ (Figure~\ref{fig:setting-Q} in the Supplementary
Material). For two-sample designs with fixed total labeled and
unlabeled budgets, balanced allocation remains best and the analytical
curves continue to track the empirical results closely across imbalance
ratios (Figure~\ref{fig:setting-R} in the Supplementary Material).

Finally, we rerun the one-sample mean experiments under two
non-Gaussian outcome distributions on the same $(n, N, \rho)$ grid. For
$t_5$ outcomes, agreement with the Gaussian-based formula remains tight,
with a maximum discrepancy of $0.02$. For log-normal outcomes, the
approximation remains reasonable once $n$ is moderate, but the most
skewed, small-$n$ configurations show visibly larger departures, with
a maximum discrepancy of $0.14$. This suggests that moderate heavy
tails alone are less problematic, whereas small samples combined with
strong skewness can pose challenges for accurate power planning.

\section{Real Data Applications}
\label{sec:applications}
In this section, we illustrate the power-analysis workflow on three
biomedical datasets: cell-type classification in single-cell
RNA-seq~\citep{baron2016single}, blood pressure estimation in NHANES
2017--2018~\citep{nhanes2020}, and melanoma detection in ISIC 2020
dermoscopy images~\citep{rotemberg2021patient}. For each dataset, we
split the observed data into training, pilot, and evaluation subsets.
The prediction models $f$ are fit on the training split to mimic a
pretrained model; the pilot split provides the variance and covariance
inputs used to plan the labeled sample size $n^\star$; and the
evaluation split is treated as an empirical population with full-data
truth $\theta_{\mathrm{eval}}$. Within each application, we specify a
target effect size $\Delta$ and set the null value to
$\theta_0 = \theta_{\mathrm{eval}} - \Delta$, so the held-out
evaluation split serves as the alternative truth. In all three
applications we allocate $25\%$ of the observed data to training,
$15\%$ to the pilot split, and the remaining $60\%$ to held-out
evaluation. This keeps the pilot meaningfully smaller than the analysis
population while still leaving enough data to estimate the planning
inputs. For ISIC, where multiple images come from the same patient,
this split is done at the patient level to avoid leakage across
repeated lesions. Throughout this section, all designs target power
$0.80$ at level $\alpha = 0.05$. Achieved power is estimated
empirically from $500$ held-out resamples: for the classical design we
repeatedly draw $n_{\mathrm{cl}}$ labeled observations from the
evaluation split, while for \texttt{PPI++} we draw $n^\star$ labeled
observations together with $N$ additional unlabeled observations from
the remaining held-out units, apply the corresponding test of
$H_0: \theta = \theta_0$, and record the rejection rate.

\subsection{Cell-Type Classification in Single-Cell RNA-Seq}

\citet{baron2016single} constructed a pancreas scRNA-seq atlas that profiles pancreatic cell types across donors. We focus on a simple composition question: among alpha and beta cells, how different
are their population shares? Restricting attention to those two cell
types, the target estimand is
$\theta = p_{\beta} - p_{\alpha}$, the difference in cell-type
proportions. The covariates $X$ are the cell-level gene-expression
profiles used to predict cell identity. We split the $4{,}851$
beta-or-alpha cells into $1{,}212$ training cells, $727$ pilot cells,
and $2{,}912$ evaluation cells. In the evaluation split,
$p_{\beta} = 0.521$ and $p_{\alpha} = 0.479$, giving
$\theta_{\mathrm{eval}} = 0.041$.

To create a meaningful gradient of prediction quality, we compare three
prediction rules trained on the training split: a deliberately weak
donor-only logistic model that uses only donor identity, a one-gene
logistic model using \texttt{INS} expression from the Baron data
itself, and a random forest fit on the top $20$ marker genes ranked
within the training split by differential mean log-expression between
beta and alpha cells rather than imported from an external marker
database.
The pilot split is used to estimate the corresponding
$\rho_{Yf}^2$ values and to plan the labeled sample size for
$\Delta = 0.10$ and $N \in \{500, 1{,}000, 2{,}000\}$.

The contrast between weak and strong predictions is immediate. At
$N = 500$, the classical design requires $n_{\mathrm{cl}} = 785$
labeled cells. The donor-only surrogate has $\rho_{Yf}^2 = 0.14$ and
yields only a mild reduction to $n^\star = 740$ with achieved power
$0.83$. By contrast, the \texttt{INS} surrogate has $\rho_{Yf}^2 =
0.97$ and reduces the plan to $n^\star = 325$, while the $20$-marker
random forest reaches $\rho_{Yf}^2 = 0.98$ and reduces it further to
$n^\star = 306$, with held-out powers $0.83$ and $0.84$. These
patterns are summarized in Figure~\ref{fig:applications-summary}.

\PPIApplicationsFigurePlacement

\subsection{Blood Pressure Estimation from NHANES}

Our second case study considers systolic blood pressure in the
2017--2018 NHANES data~\citep{nhanes2020}. We merge the demographic,
anthropometric, and blood-pressure examination files and retain
$4{,}990$ adults with complete covariates and at least one systolic
blood-pressure measurement. Here, $Y$ is the mean systolic blood pressure
and $X = (\text{age, sex, race, BMI, waist circumference})$. The
complete dataset is split into $1{,}247$ training observations,
$748$ pilot observations, and $2{,}995$ evaluation observations; the
evaluation mean is $\theta_{\mathrm{eval}} = 125.9$~mmHg.

We compare three predictive models fit on the training split: an age-only
linear model, a richer clinical linear model using all available
covariates, and a random forest on all available covariates.
The resulting gains are more moderate. At
$N = 2{,}000$, the classical design requires $n_{\mathrm{cl}} = 187$
labeled patients. The age-only model has $\rho_{Yf}^2 = 0.22$ and
reduces the plan to $n^\star = 149$ with held-out power $0.84$. The
richer clinical linear model has $\rho_{Yf}^2 = 0.24$ and yields
$n^\star = 144$ with held-out power $0.85$, while the clinical random
forest has $\rho_{Yf}^2 = 0.26$ and yields $n^\star = 142$ with
held-out power $0.81$. This case study shows more incremental
gains, around 20\%, and
the smaller pilot split leaves more planning noise in held-out
validation.

\subsection{Melanoma Detection in Dermoscopy Images}
Given the global shortage of dermatology resources, screening for skin
cancer---especially melanoma---remains time-consuming, expensive, and
difficult to scale, contributing to delayed diagnoses and rising
treatment burdens. Recent work has shown much promise in automated
dermoscopy classifiers for melanoma
detection~\citep{daneshjou2022disparities,xu2024foundation,chen2024singlelesion}. For this case study, we use the ISIC 2020 challenge
data~\citep{rotemberg2021patient}, which contains $33{,}126$
dermoscopy images from $2{,}056$ patients with a melanoma prevalence
of $1.8\%$.  The target estimand is the mean prevalence.  We use the
$256$-pixel thumbnails of dermoscopy images as well as the
accompanying metadata (e.g., age, sex, anatomic site, and patient
lesion counts) for a non-image baseline.  We split the data at the
patient level into $8{,}609$ training images, $5{,}090$ pilot images,
and $19{,}427$ evaluation images, corresponding to $514$, $308$, and
$1{,}234$ patients, respectively.  The evaluation prevalence is
$\theta_{\mathrm{eval}} = 0.018$.

We compare three predictors $f$ trained on the training split: a
metadata-only logistic baseline, a logistic model on the top principal
components of the image thumbnails, and a logistic regression on the CLIP
embeddings~\citep{radford2021clip} of the image, which have been shown
to perform well in melanoma detection~\citep{xu2024foundation}. The pilot split is then used to
plan the labeled sample size needed to detect an absolute prevalence
shift of $\Delta = 0.01$ with
$N \in \{1{,}000, 5{,}000, 10{,}000\}$ unlabeled images.  On the
held-out evaluation split, these three models achieve AUROC values of
$0.72$, $0.79$, and $0.83$, respectively.

Despite the reasonable AUROCs, this is an intrinsically hard setting
with $1.8\%$ population prevalence.  The resulting gains are therefore minimal.  At $N = 10{,}000$, the classical design
requires $n_{\mathrm{cl}} = 1{,}334$ pathology-confirmed labels.  The
metadata-only surrogate yields $n^\star = 1{,}299$ with held-out
power $0.868$, the thumbnail-PCA surrogate yields
$n^\star = 1{,}310$ with held-out power $0.872$, and the CLIP
surrogate yields $n^\star = 1{,}296$ with held-out power $0.832$.
Unlike Baron and NHANES, the ISIC case study stays close to the
classical regime: even visibly stronger discrimination in the image
models does not translate into large planning gains when the outcome
is this rare, as the $\rho_{Yf}^2$ remains very modest ($<0.05$ in
all cases).

\paragraph{Takeaway.}
These three examples make the role of the predictive model concrete.
In the Baron data, moving from donor-only metadata to expression-based
classifiers shifts the design from nearly classical to sharply
prediction-powered.  In NHANES, all three models lie in a
moderate-$R^2$ regime, so the gains are real but modest.  ISIC shows a
third regime: even when held-out AUROC improves visibly from metadata
to image-based models, the rarity of melanoma leaves the design close
to classical. In short, prediction-powered inference is not magic:
sample-size reduction is determined by the agreement between
predictions and true labels, as measured here by correlation-based
summaries.

\section{Discussion}
\label{sec:discussion}

In this paper, we tackle a practical design question in the era of AI/ML: given a predictive model's
accuracy, how many fewer samples can a study use by leveraging predictions for inference?
For the one-sample mean, the rule-of-thumb reduction is around $R^2$---a model explaining half of the outcome variation roughly halves the required sample size. Through extensive simulations, we found that our proposed formulas yield theoretical power and sample sizes that closely match the empirical power and sample sizes.
Noticeable departures occur at very small sample sizes (fewer than 20 labeled samples) due to high sampling variability.
We demonstrate the plug-and-play use of the method on three real-world datasets and open-source our software.
Software implementing the formulas is available in the
\href{https://github.com/yiqunchen/pppower}{GitHub repository} and in browser-based
\href{https://yiqunchen.github.io/pppower/articles/sample-size-calculator.html}{sample-size}
and
\href{https://yiqunchen.github.io/pppower/articles/calculator-2x2.html}{$2\times 2$}
planning calculators.

There are several natural extensions of our work. While our current framework largely relies on asymptotic normality, the power analysis literature provides a rich set of small-sample variance and distributional corrections that could improve finite-sample approximations.
Similarly, we largely considered the predictor $f$ to be either pretrained (e.g., ChatGPT or CLIP) or simple enough for cross-fitting to work reliably.
A more extensive set of simulations examining finite-sample variance could provide a more precise characterization of these effects. Finally, the prediction-powered inference framework has mostly focused on independent and identically distributed data. Extending the theory, as well as practical sample size design, to more realistic settings with complex designs such as clustered, longitudinal, or survival data is an important next step.

\section*{Acknowledgements}
The authors declare no conflict of interest. This work was partially supported by the Johns Hopkins Bloomberg School of Public Health, Department of Biostatistics, Data Science and AI Faculty Innovation Fund and a Google Academic Research Award on AI for Privacy, Safety, and Security.

\section*{Data Availability}
Software implementing all methods is available as an \texttt{R} package at \url{https://github.com/yiqunchen/pppower}. The datasets used in the case studies are publicly available from their original publications: the Baron scRNA-seq data~\citep{baron2016single}, the NHANES 2017--2018 data~\citep{nhanes2020}, and the ISIC 2020 dermoscopy data~\citep{rotemberg2021patient}.

\PPIBibliographyBlock

\PPIFiguresAtEnd

\clearpage
\thispagestyle{empty}
\begin{center}
{\fontsize{16}{19}\selectfont Supplementary Material}\\[1.5em]
{\fontsize{16}{19}\selectfont for}\\[1.5em]
{\fontsize{16}{19}\selectfont Power Analysis for Prediction-Powered Inference}
\end{center}

\vspace{1.5em}
\appendix
\setcounter{table}{0}
\setcounter{figure}{0}
\renewcommand{\thetable}{S\arabic{table}}
\renewcommand{\thefigure}{S\arabic{figure}}
\renewcommand{\theHtable}{supp.\arabic{table}}
\renewcommand{\theHfigure}{supp.\arabic{figure}}
\section{Supplementary Material}
\label{sec:appendix}

\subsection{Proofs for Sections~\ref{sec:ppi-vanilla}--\ref{sec:ttest}}
\label{sec:proofs-main}

We collect the proofs for the one-sample, two-sample, paired, and
$2\times 2$ results stated in the main text, together with the
cross-fitting extension. For clarity, each proof
subsection is labeled by the result it proves. The regression-contrast
derivations and the proof of
Proposition~\ref{prop:regression-n} appear in
Appendix~\ref{sec:regression-appendix}.

\input{proofs}
\end{document}

%% file: proofs.tex
\setstretch{2} 


\subsubsection{One-Sample Mean Results}
\label{sec:proofs-one-sample}

\paragraph{Proof of Proposition~\ref{prop:ppi-var} (Mean PPI/PPI++ Variance)}
\label{sec:proof-vanilla-var}
This proof covers both the PPI and \texttt{PPI++} variance results stated in Proposition~\ref{prop:ppi-var}.

Write
\[
\wh{\theta}_{\mathrm{PPI}}
= \bar f_U + (\bar Y_L-\bar f_L),
\qquad
\wh{\theta}_\lambda
= \bar Y_L + \lambda(\bar f_U-\bar f_L),
\]
where $\bar f_U = N^{-1}\sum_{j=1}^N \wt f_j$,
$\bar f_L = n^{-1}\sum_{i=1}^n f_i$, and
$\bar Y_L = n^{-1}\sum_{i=1}^n Y_i$.
Because the labeled and unlabeled splits are independent and drawn from the
same covariate distribution,
\[
\E[\bar f_U]=\E[\bar f_L]=\E[f(X)],
\]
so
\[
\E[\wh{\theta}_{\mathrm{PPI}}]
= \E[f(X)] + \E[Y-f(X)]
= \theta^\star,
\qquad
\E[\wh{\theta}_\lambda]=\theta^\star
\]
for every fixed $\lambda$.

For vanilla PPI, independence of the labeled and unlabeled terms gives
\[
\Var(\wh{\theta}_{\mathrm{PPI}})
= \Var(\bar f_U) + \Var(\bar Y_L-\bar f_L)
= \frac{\sigma_f^2}{N} + \frac{\sigma_\varepsilon^2}{n},
\]
which is~\eqref{eq:ppi-var}.

For \texttt{PPI++},
\[
\Var(\wh{\theta}_\lambda)
= \Var(\bar Y_L) + \lambda^2\Var(\bar f_U-\bar f_L)
+ 2\lambda\,\Cov(\bar Y_L,\bar f_U-\bar f_L).
\]
Now $\Var(\bar Y_L)=\sigma_Y^2/n$,
$\Var(\bar f_U-\bar f_L)=\sigma_f^2/N+\sigma_f^2/n$, and
$\Cov(\bar Y_L,\bar f_U)=0$ by split independence, while
$\Cov(\bar Y_L,\bar f_L)=\Cov(Y,f)/n$.
Substituting these identities yields
\[
\Var(\wh{\theta}_\lambda)
=
\frac{\sigma_Y^2}{n}
+ \lambda^2\!\left(\frac{\sigma_f^2}{N}+\frac{\sigma_f^2}{n}\right)
- \frac{2\lambda\,\Cov(Y,f)}{n},
\]
which is~\eqref{eq:ppi++-variance}. Joint asymptotic normality follows from the
multivariate central limit theorem applied to the labeled and unlabeled sample
means.

\paragraph{Justification of the Classical Power Approximation}
\label{sec:proof-vanilla-power}
Under the alternative $\theta^\star=\theta_0+\Delta$, the classical Wald
statistic is asymptotically normal with mean
$\Delta/(\sigma_Y/\sqrt{n})$ and variance $1$, so the exact two-sided normal
approximation is~\eqref{eq:classical-power}. The second term in that display,
\[
\Phi\!\left(-z_{1-\alpha/2}-\frac{|\Delta|}{\sigma_Y/\sqrt{n}}\right),
\]
is always no larger than the first and becomes negligible once the noncentrality
$|\Delta|/(\sigma_Y/\sqrt{n})$ is moderate. Dropping it yields the usual
planning approximation
$|\Delta|/(\sigma_Y/\sqrt{n}) \ge z_{1-\alpha/2}+z_{1-\beta}$, or equivalently
$\sigma_Y^2/n \le S^2$, which is the inversion used in the main text.

\paragraph{Proof of Equations~\eqref{eq:lambda-star} and~\eqref{eq:ppi++-var-opt}}
Given~\eqref{eq:ppi++-variance},
\[
g(\lambda)
= \Var(\wh{\theta}_\lambda)
= \frac{\sigma_Y^2}{n}
+ \lambda^2\!\left(\frac{\sigma_f^2}{N}+\frac{\sigma_f^2}{n}\right)
- \frac{2\lambda\,\Cov(Y,f)}{n}.
\]
This is a convex quadratic in $\lambda$, so its minimizer satisfies
\[
g'(\lambda)
= 2\lambda\!\left(\frac{\sigma_f^2}{N}+\frac{\sigma_f^2}{n}\right)
- \frac{2\Cov(Y,f)}{n}
= 0.
\]
Solving gives
\[
\lambda^\star
= \frac{\Cov(Y,f)}{\sigma_f^2(1+n/N)}
= \frac{\Cov(Y,f)}{(1+r)\sigma_f^2},
\]
which is~\eqref{eq:lambda-star}. Substituting this value back into
\eqref{eq:ppi++-variance} yields
\[
\Var(\wh{\theta}_{\lambda^\star})
= \frac{\sigma_Y^2}{n}
- \frac{\Cov(Y,f)^2}{\sigma_f^2}\cdot \frac{N}{n(n+N)},
\]
which is~\eqref{eq:ppi++-var-opt}.

\paragraph{Proof of Proposition~\ref{prop:ppi++-power} (\texttt{PPI++} Power)}
Under $\theta^\star=\theta_0+\Delta$,
\[
Z
= \frac{\wh{\theta}_{\lambda^\star}-\theta_0}
{\sqrt{\Var(\wh{\theta}_{\lambda^\star})}}
\;\dot{\sim}\;
\mathcal N\!\left(
\frac{\Delta}{\sqrt{\Var(\wh{\theta}_{\lambda^\star})}},\,1
\right),
\]
so the two-sided rejection probability is
\[
P(|Z|\ge z_{1-\alpha/2})
=
\Phi\!\left(-z_{1-\alpha/2}
+ \frac{|\Delta|}{\sqrt{\Var(\wh{\theta}_{\lambda^\star})}}\right)
+ \Phi\!\left(-z_{1-\alpha/2}
- \frac{|\Delta|}{\sqrt{\Var(\wh{\theta}_{\lambda^\star})}}\right),
\]
which is exactly~\eqref{eq:ppi++-power}.

\paragraph{Proof of Proposition~\ref{prop:ppi++-n} (Sample Size Inversion)}
\label{sec:proof-ppi++-n}
From Proposition~\ref{prop:ppi++-power}, achieving target power is ensured by
\[
\Var(\wh{\theta}_{\lambda^\star}) \le S^2,
\qquad
S^2 = \frac{\Delta^2}{(z_{1-\alpha/2}+z_{1-\beta})^2}.
\]
Substituting~\eqref{eq:ppi++-var-opt} gives
\[
\frac{\sigma_Y^2}{n}
- \frac{\Cov(Y,f)^2}{\sigma_f^2}\cdot\frac{N}{n(n+N)}
\le S^2.
\]
Multiplying by $n(n+N)>0$ and using
$\rho_{Yf}^2=\Cov(Y,f)^2/(\sigma_Y^2\sigma_f^2)$ yields
\[
S^2 n^2 + (S^2N-\sigma_Y^2)n - N\sigma_Y^2(1-\rho_{Yf}^2)\ge 0.
\]
Since this quadratic opens upward, the feasible region is $n$ greater than or
equal to its positive root:
\[
n \ge
\frac{\sigma_Y^2 - S^2 N
+ \sqrt{(\sigma_Y^2 - S^2 N)^2
+ 4 S^2 N \sigma_Y^2 (1-\rho_{Yf}^2)}}{2S^2},
\]
which is~\eqref{eq:ppi++-n-formula}.

\paragraph{Proof of Corollary~\ref{cor:rule-of-thumb} (Rule of Thumb)}
When $N \gg n$, we have $N/(n+N)=1+o(1)$ and therefore
\[
\Var(\wh{\theta}_{\lambda^\star})
= \frac{\sigma_Y^2}{n}
- \frac{\Cov(Y,f)^2}{\sigma_f^2}\cdot\frac{1+o(1)}{n}
= \frac{\sigma_Y^2(1-\rho_{Yf}^2)}{n} + o(n^{-1}).
\]
The classical variance is $\sigma_Y^2/n$, so to first order the required
labeled sample size scales by the same factor, giving
$n_{\mathrm{PPI}}/n_{\mathrm{cl}} \approx 1-\rho_{Yf}^2$.

\subsubsection{Two-Sample, Paired, and Contingency-Table Extensions}
\label{sec:proofs-extensions}

\paragraph{Proof of Proposition~\ref{prop:ttest-var} (Two-Sample Variance)}
\label{sec:proof-ttest}
From the definition~\eqref{eq:ppi++-ttest-est},
\[
\wh{\Delta}_{\mathrm{PPI}}
= \wh{\theta}_{A,\mathrm{PPI}}-\wh{\theta}_{B,\mathrm{PPI}}.
\]
Applying the one-sample oracle-variance formula~\eqref{eq:ppi++-var-opt}
within each group gives
\[
\Var(\wh{\theta}_{g,\lambda_g^\star})
=
\frac{\sigma_{Y,g}^2}{n_g}
- \frac{\Cov(Y_g,f_g)^2}{\sigma_{f,g}^2}
\cdot \frac{N_g}{n_g(n_g+N_g)},
\qquad g\in\{A,B\}.
\]
Because the two groups are independent, the cross-covariance vanishes, so
\[
\Var(\wh{\Delta}_{\mathrm{PPI}})
= \Var(\wh{\theta}_{A,\lambda_A^\star})
+ \Var(\wh{\theta}_{B,\lambda_B^\star}),
\]
which is exactly~\eqref{eq:ppi++-ttest-var}. The approximation
\eqref{eq:ppi++-ttest-var-approx} follows from
$\Cov(Y_g,f_g)^2/\sigma_{f,g}^2 = \sigma_{Y,g}^2\Corr(Y_g,f_g)^2$ and
$N_g/(n_g+N_g)=1+O(n_g/N_g)$.

\paragraph{Proof of Proposition~\ref{prop:paired-var} (Paired Variance)}
Let $D=Y^A-Y^B$ and $G=f^A-f^B$. The paired \texttt{PPI++} estimator is exactly
the one-sample \texttt{PPI++} estimator applied to the paired differences, with
target parameter $\E[D]$ and surrogate $G$. Proposition~\ref{prop:ppi-var}
therefore gives
\[
\Var(\wh{\Delta}_{\mathrm{PPI++}})
= \frac{\Var(D)}{n}
- \frac{\Cov(D,G)^2}{\Var(G)}\cdot \frac{N}{n(n+N)}.
\]
Rewriting $\Cov(D,G)^2/\Var(G)=\Var(D)\Corr(D,G)^2$ gives the equivalent form in
the proposition, and letting $N\gg n$ yields the displayed approximation.

\paragraph{Proof of Proposition~\ref{prop:2x2-n} ($2\times2$ Sample Size)}
For each group $g\in\{0,1\}$, the event-rate estimator
\[
\wh p_{g,\lambda_g}
= \bar Y_{L,g} + \lambda_g(\bar f_{U,g}-\bar f_{L,g})
\]
is a one-sample \texttt{PPI++} estimator for the binary mean $p_g$. Since
$Y_g\in\{0,1\}$, $\Var(Y_g)=p_g(1-p_g)$, and in the regime $N_g\gg n_g$,
Corollary~\ref{cor:rule-of-thumb} gives
\[
\Var(\wh p_{g,\lambda_g^\star})
\approx \frac{p_g(1-p_g)(1-\rho_g^2)}{n_g}.
\]

For $\log(\mathrm{RR})=\log p_1-\log p_0$, the gradient with respect to
$(p_0,p_1)$ is $(-1/p_0,\;1/p_1)^\top$. Since the two groups are independent,
the delta method yields
\[
\Var(\log \widehat{\mathrm{RR}})
\approx
\frac{\Var(\wh p_{0,\lambda_0^\star})}{p_0^2}
+ \frac{\Var(\wh p_{1,\lambda_1^\star})}{p_1^2}
=
\frac{1}{n_0}\frac{(1-p_0)(1-\rho_0^2)}{p_0}
+ \frac{1}{n_1}\frac{(1-p_1)(1-\rho_1^2)}{p_1}.
\]
Substituting $n_1=\kappa n_0$ gives the first displayed variance formula in the
proposition.

For $\log(\mathrm{OR})=\operatorname{logit}(p_1)-\operatorname{logit}(p_0)$, the
gradient is
$(-[p_0(1-p_0)]^{-1},\; [p_1(1-p_1)]^{-1})^\top$, so the same argument gives
\[
\Var(\log \widehat{\mathrm{OR}})
\approx
\frac{1}{n_0}\frac{1-\rho_0^2}{p_0(1-p_0)}
+ \frac{1}{n_1}\frac{1-\rho_1^2}{p_1(1-p_1)}.
\]
The sample-size expressions follow by imposing
$\Var(\log \widehat{\mathrm{RR}})\le S_{\mathrm{RR}}^2$ and
$\Var(\log \widehat{\mathrm{OR}})\le S_{\mathrm{OR}}^2$ and then solving for
$n_0$; the balanced-design formulas set $\kappa=1$. Setting
$\rho_0=\rho_1=0$ recovers the usual classical formulas.

\subsubsection{Cross-Fitted PPI++ Distribution}
\label{sec:proof-cross-fitting}

We state the cross-fitting result directly for the one-sample mean problem of
Sections~\ref{sec:ppi-vanilla}--\ref{sec:ppi++}. Partition the labeled sample
into folds $I_1,\ldots,I_K$, let $f^{(j)}$ denote the predictor trained
without fold $I_j$, and define the cross-fitted \texttt{PPI++} estimator
\[
\wh{\theta}_{\lambda,\mathrm{cf}}
=
\bar Y_L
+
\lambda\left(
\frac{1}{K}\sum_{j=1}^K \bar f_{U}^{(j)}
-
\frac{1}{n}\sum_{j=1}^K\sum_{i\in I_j} f^{(j)}(X_i)
\right),
\qquad
\bar f_U^{(j)} = \frac{1}{N}\sum_{m=1}^N f^{(j)}(\wt X_m).
\]
This is the same rectified mean estimator as in~\eqref{eq:ppi++-estimator},
except that each labeled observation is evaluated by a predictor that was not
trained on its own fold.

\begin{assumption}[Prediction stability]\label{ass:M-stable-f}
The predictors $f^{(1)},\ldots,f^{(K)}$ trained on the $K$ folds satisfy
$\|f^{(j)} - \bar f\|_2 = o_P(1)$ as $n \to \infty$,
where $\bar f = K^{-1}\sum_{j=1}^K f^{(j)}$ is the average predictor.
\end{assumption}

Lemmas~3--4 in the proof of Theorem~2 of~\citet{zrnic2024cross}, specialized
to the scalar mean estimating function $\ell_\theta(Y,X)=Y-\theta$, imply
that the fold-specific prediction averages can be replaced by the average
predictor $\bar f$ without affecting the estimator at the $n^{-1/2}$ scale:
\begin{align*}
\frac{1}{K}\sum_{j=1}^K \bar f_U^{(j)}
&=
\frac{1}{N}\sum_{m=1}^N \bar f(\wt X_m) + o_P(n^{-1/2}), \\
\frac{1}{n}\sum_{j=1}^K\sum_{i\in I_j} f^{(j)}(X_i)
&=
\frac{1}{n}\sum_{i=1}^n \bar f(X_i) + o_P(n^{-1/2}).
\end{align*}
Therefore
\[
\wh{\theta}_{\lambda,\mathrm{cf}}
=
\bar Y_L + \lambda(\bar f_{U,\bar f} - \bar f_{L,\bar f}) + o_P(n^{-1/2}),
\]
where
\[
\bar f_{U,\bar f} = \frac{1}{N}\sum_{m=1}^N \bar f(\wt X_m),
\qquad
\bar f_{L,\bar f} = \frac{1}{n}\sum_{i=1}^n \bar f(X_i).
\]

Conditional on the limiting predictor $\bar f$, this is exactly the same
algebraic form as the fixed-predictor estimator in
\eqref{eq:ppi++-estimator}, with $f$ replaced by $\bar f$. Since the labeled
and unlabeled samples are independent, the multivariate central limit theorem
gives
\[
\sqrt{n}\bigl(\wh{\theta}_{\lambda,\mathrm{cf}} - \theta^\star\bigr)
\overset{d}{\longrightarrow}
\mathcal N\!\left(
0,\;
\sigma_Y^2 + \lambda^2(1+r)\sigma_{\bar f}^2 - 2\lambda\,\Cov(Y,\bar f)
\right),
\]
where $r = \lim n/N$ and $\sigma_{\bar f}^2 = \Var(\bar f(X))$. Equivalently,
\[
\Var(\wh{\theta}_{\lambda,\mathrm{cf}})
=
\frac{\sigma_Y^2}{n}
+
\lambda^2\sigma_{\bar f}^2\!\left(\frac{1}{n}+\frac{1}{N}\right)
-
\frac{2\lambda\,\Cov(Y,\bar f)}{n}
+
o(n^{-1}).
\]
The oracle choice in this cross-fitted setting is therefore
\[
\lambda_{\mathrm{cf}}^\star
=
\frac{\Cov(Y,\bar f)}{(1+r)\sigma_{\bar f}^2},
\]
and substituting it into the variance expression yields the same first-order
formula as \eqref{eq:ppi++-var-opt}, again with $f$ replaced by the limiting
predictor $\bar f$.

\subsection{Supplementary Simulation Results}
\label{sec:appendix-sim-summary}

\paragraph{Head-to-head planning comparison.}
Table~\ref{tab:head-to-head-multidesign} reorganizes representative
results from the main simulation settings into a single comparison
across one-sample, two-sample, and paired continuous designs.  It does
not introduce a new setting; rather, it summarizes the practical
tradeoffs between classical inference, vanilla PPI, and
\texttt{PPI++} (oracle and plugin) using common planning targets.

\begin{table}[!ht]
\centering
\caption{Head-to-head comparison across representative continuous
  planning scenarios (one-sample: $\Delta = 0.2$; two-sample:
  $\Delta = 0.3$; paired: $\Delta = 0.3$; all with $N = 5000$,
  $\rho = 0.7$, target power~$0.80$, $\alpha = 0.05$;
  MC $R = 5000$).  For plugin \texttt{PPI++}, planning uses the oracle
  closed-form $n$ and evaluates performance with
  plugin~$\hat\lambda$.}
\label{tab:head-to-head-multidesign}
\small
\begin{tabular}{llrrrr}
\toprule
Design & Method & Required $n$ & Achieved power & Type~I error
  & Reduction \\
\midrule
One-sample & Classical & 197 & 0.797 & 0.048 & 0.0\% \\
One-sample & Vanilla PPI & 123 & 0.801 & 0.052 & 37.6\% \\
One-sample & \texttt{PPI++} (Oracle) & 102 & 0.811 & 0.048 & 48.2\% \\
One-sample & \texttt{PPI++} (Plugin) & 102 & 0.804 & 0.056 & 48.2\% \\
Two-sample & Classical & 175 & 0.796 & 0.046 & 0.0\% \\
Two-sample & Vanilla PPI & 109 & 0.791 & 0.048 & 37.7\% \\
Two-sample & \texttt{PPI++} (Oracle) & 91 & 0.800 & 0.054 & 48.0\% \\
Two-sample & \texttt{PPI++} (Plugin) & 91 & 0.808 & 0.050 & 48.0\% \\
Paired & Classical & 88 & 0.801 & 0.050 & 0.0\% \\
Paired & Vanilla PPI & 54 & 0.810 & 0.054 & 38.6\% \\
Paired & \texttt{PPI++} (Oracle) & 45 & 0.802 & 0.049 & 48.9\% \\
Paired & \texttt{PPI++} (Plugin) & 45 & 0.805 & 0.055 & 48.9\% \\
\bottomrule
\end{tabular}
\end{table}

\subsection{Additional Application Summary}
\label{sec:appendix-application-summary}

Table~\ref{tab:applications} collects representative design points from
the three real-data applications in Section~\ref{sec:applications}.

\begin{table}[!ht]
\centering
\caption{Representative planning outputs and held-out achieved power
  from the $15\%$ pilot / $60\%$ held-out workflow.  The Baron row
  uses the beta-minus-alpha composition contrast at $N = 500$; NHANES
  uses mean systolic blood pressure at $N = 2{,}000$; and ISIC uses
  melanoma prevalence at $N = 10{,}000$.}
\label{tab:applications}
\small
\begin{tabular}{llccrrr}
\toprule
Application & Surrogate & $\rho_{Yf}^2$ & $\Delta$ & $n^\star$
  & Achieved power & Reduction \\
\midrule
Baron & Classical & --- & 0.10 & 785 & 0.840 & --- \\
Baron & Donor GLM & 0.141 & 0.10 & 740 & 0.826 & 6\% \\
Baron & INS GLM & 0.967 & 0.10 & 325 & 0.826 & 59\% \\
Baron & Top-20 RF & 0.983 & 0.10 & 306 & 0.840 & 61\% \\
NHANES & Classical & --- & 4~mmHg & 187 & 0.830 & --- \\
NHANES & Age-only LM & 0.218 & 4~mmHg & 149 & 0.836 & 20\% \\
NHANES & Clinical LM & 0.244 & 4~mmHg & 144 & 0.848 & 23\% \\
NHANES & Clinical RF & 0.258 & 4~mmHg & 142 & 0.812 & 24\% \\
ISIC & Classical & --- & 0.01 & 1{,}334 & 0.806 & --- \\
ISIC & Metadata GLM & 0.030 & 0.01 & 1{,}299 & 0.868 & 3\% \\
ISIC & Thumbnail PCA GLM & 0.020 & 0.01 & 1{,}310 & 0.872 & 2\% \\
ISIC & CLIP GLM & 0.032 & 0.01 & 1{,}296 & 0.832 & 3\% \\
\bottomrule
\end{tabular}
\end{table}

\subsection{Regression and GLM Extensions}
\label{sec:regression-appendix}

This single appendix subsection collects the OLS and GLM contrast
derivations together with the proof of
Proposition~\ref{prop:regression-n}.  It supports the main-text
regression subsection rather than introducing a separate appendix
module, and uses the same contrast-level notation
$(V_{YY}, V_{ff}, V_{Yf})$ as the main text. These derivations are
validated by the Monte Carlo regression-contrast experiments in
Section~\ref{sec:sim-regression}.

\paragraph{Linear Regression (OLS).}
Consider the population least-squares target
\[
\beta^\star
=
\arg\min_{\beta\in\mathbb R^p}
\E\big[(Y-X^\top\beta)^2\big].
\]
Its first-order condition is
\[
\E\big[X\{Y-X^\top\beta^\star\}\big]=0,
\]
or equivalently
\[
\E[XX^\top]\beta^\star=\E[XY].
\]
Let
\[
\Sigma_{XX}:=\E[XX^\top].
\]
When $\Sigma_{XX}$ is nonsingular, $\beta^\star=\Sigma_{XX}^{-1}\E[XY]$.

With an external predictor $f(X)$, we can decompose
\[
\E[XY]
=
\E[Xf(X)] + \E[X\{Y-f(X)\}],
\]
which leads to the empirical PPI estimator
\[
\hat\beta_{\mathrm{PPI}}
=
\hat\Sigma_{XX,U}^{-1}
\left[
\frac{1}{N}\sum_{j=1}^N \tilde X_j\tilde f_j
+
\frac{1}{n}\sum_{i=1}^n X_i\{Y_i-f_i\}
\right],
\]
where
\[
\hat\Sigma_{XX,U}=\frac{1}{N}\sum_{j=1}^N \tilde X_j\tilde X_j^\top.
\]

For a contrast $\theta=a^\top\beta$, define the OLS score residuals
\[
\psi_Y(X,Y)=X\{Y-X^\top\beta^\star\},
\qquad
\psi_f(X)=X\{f(X)-X^\top\beta^\star\},
\]
and the corresponding covariance blocks
\[
\Sigma_{YY}=\Var\{\psi_Y(X,Y)\},
\qquad
\Sigma_{ff}=\Var\{\psi_f(X)\},
\qquad
\Sigma_{Yf}=\Cov\{\psi_Y(X,Y),\psi_f(X)\}.
\]
The contrast-level quantities used in the main text are then
\begin{align*}
V_{YY}^{\mathrm{OLS}}
&=
a^\top\Sigma_{XX}^{-1}\Sigma_{YY}\Sigma_{XX}^{-1}a, \\
V_{ff}^{\mathrm{OLS}}
&=
a^\top\Sigma_{XX}^{-1}\Sigma_{ff}\Sigma_{XX}^{-1}a, \\
V_{Yf}^{\mathrm{OLS}}
&=
a^\top\Sigma_{XX}^{-1}\Sigma_{Yf}\Sigma_{XX}^{-1}a.
\end{align*}

The rectified \texttt{PPI++} estimator solves
\[
0
=
\frac{1}{n}\sum_{i=1}^n X_i\big(Y_i-X_i^\top\beta\big)
+
\lambda\left[
\frac{1}{N}\sum_{j=1}^N \tilde X_j\big(\tilde f_j-\tilde X_j^\top\beta\big)
-
\frac{1}{n}\sum_{i=1}^n X_i\big(f_i-X_i^\top\beta\big)
\right].
\]
A first-order expansion around $\beta^\star$ gives
\[
\Var\!\big(a^\top\hat\beta_\lambda\big)
\approx
\frac{
V_{YY}^{\mathrm{OLS}}
+
\lambda^2(1+r)V_{ff}^{\mathrm{OLS}}
-2\lambda V_{Yf}^{\mathrm{OLS}}
}{n},
\qquad
r=\frac{n}{N}.
\]
The oracle tuning parameter is therefore
$\lambda^\star(a)=V_{Yf}^{\mathrm{OLS}}/\{(1+r)V_{ff}^{\mathrm{OLS}}\}$,
and the corresponding fixed-$N$ sample-size formula is exactly the one
stated in Proposition~\ref{prop:regression-n}; Appendix~\ref{sec:proof-regression-n}
contains the algebraic inversion.

\paragraph{Generalized Linear Models (GLMs).}
For a GLM with mean function $\mu_\beta(X)$, the target parameter
$\beta^\star$ satisfies the score equation
\[
\E\big[X\{Y-\mu_\beta(X)\}\big]=0
\qquad\text{at }\beta=\beta^\star.
\]
If $\mu_f(X)$ is an external prediction for $\E[Y\mid X]$, then
\[
\E\big[X\{Y-\mu_\beta(X)\}\big]
=
\E\big[X\{\mu_f(X)-\mu_\beta(X)\}\big]
+
\E\big[X\{Y-\mu_f(X)\}\big].
\]
This yields the empirical PPI score
\[
\hat U_{\mathrm{PPI}}(\beta)
=
\frac{1}{N}\sum_{j=1}^N \tilde X_j\big(\mu_f(\tilde X_j)-\mu_\beta(\tilde X_j)\big)
+
\frac{1}{n}\sum_{i=1}^n X_i\big(Y_i-\mu_f(X_i)\big).
\]

Let
\[
J
:=
-\E\left[\frac{\partial}{\partial\beta}
X\{Y-\mu_\beta(X)\}\Big|_{\beta=\beta^\star}\right]
\]
denote the Fisher information matrix. For canonical links,
$J=\E[w_\star(X)XX^\top]$ with the usual working weight
$w_\star(X)$ evaluated at $\beta^\star$.

Define the GLM score residuals
\[
\phi_Y(X,Y)=X\{Y-\mu_{\beta^\star}(X)\},
\qquad
\phi_f(X)=X\{\mu_f(X)-\mu_{\beta^\star}(X)\},
\]
their covariance blocks
\[
\Sigma_{YY}^{\mathrm{GLM}}=\Var\{\phi_Y(X,Y)\},
\qquad
\Sigma_{ff}^{\mathrm{GLM}}=\Var\{\phi_f(X)\},
\qquad
\Sigma_{Yf}^{\mathrm{GLM}}=\Cov\{\phi_Y(X,Y),\phi_f(X)\},
\]
and the induced contrast-level quantities
\begin{align*}
V_{YY}^{\mathrm{GLM}}
&=
a^\top J^{-1}\Sigma_{YY}^{\mathrm{GLM}}J^{-1}a, \\
V_{ff}^{\mathrm{GLM}}
&=
a^\top J^{-1}\Sigma_{ff}^{\mathrm{GLM}}J^{-1}a, \\
V_{Yf}^{\mathrm{GLM}}
&=
a^\top J^{-1}\Sigma_{Yf}^{\mathrm{GLM}}J^{-1}a.
\end{align*}

The rectified \texttt{PPI++} score is
\begin{equation}\label{eq:ppi++-glm-score}
\hat U_\lambda(\beta)
=
\frac{1}{n}\sum_{i=1}^n X_i\big(Y_i-\mu_\beta(X_i)\big)
+
\lambda\left[
\frac{1}{N}\sum_{j=1}^N \tilde X_j\big(\mu_f(\tilde X_j)-\mu_\beta(\tilde X_j)\big)
-
\frac{1}{n}\sum_{i=1}^n X_i\big(\mu_f(X_i)-\mu_\beta(X_i)\big)
\right].
\end{equation}
Solving \eqref{eq:ppi++-glm-score} gives $\hat\beta_\lambda$. A
one-step expansion yields
\[
\Var\!\big(a^\top\hat\beta_\lambda\big)
\approx
\frac{
V_{YY}^{\mathrm{GLM}}
+
\lambda^2(1+r)V_{ff}^{\mathrm{GLM}}
-2\lambda V_{Yf}^{\mathrm{GLM}}
}{n},
\qquad
r=\frac{n}{N}.
\]
Hence
$\lambda^\star(a)=V_{Yf}^{\mathrm{GLM}}/\{(1+r)V_{ff}^{\mathrm{GLM}}\}$,
and the same fixed-$N$ sample-size inversion as in
Proposition~\ref{prop:regression-n} applies with the GLM-specific
contrast-level terms. For logistic regression in particular,
$J=\E[\mu_\star(X)\{1-\mu_\star(X)\}XX^\top]$.


\paragraph{Proof of Proposition~\ref{prop:regression-n}.}
\label{sec:proof-regression-n}
The derivations in Appendix~\ref{sec:regression-appendix} show that for either
OLS or a GLM contrast,
\[
\Var(\wh{\theta}_\lambda)
\approx
\frac{V_{YY} + \lambda^2(1+r)V_{ff} - 2\lambda V_{Yf}}{n},
\qquad
r=\frac{n}{N}.
\]
This is a convex quadratic in $\lambda$, so differentiating with respect to
$\lambda$ and setting the derivative to zero gives
\[
\lambda^\star
= \frac{V_{Yf}}{(1+r)V_{ff}}.
\]
Substituting $\lambda^\star$ back into the variance expression yields
\[
\Var(\wh{\theta}_{\lambda^\star})
\approx
\frac{1}{n}\left(
V_{YY}
- \frac{V_{Yf}^2}{(1+r)V_{ff}}
\right)
=
\frac{1}{n}\left(
V_{YY}
- \frac{V_{Yf}^2}{V_{ff}}\cdot\frac{N}{N+n}
\right).
\]

For a two-sided Wald test, the same argument as in the one-sample case shows
that target power is achieved whenever
\[
\Var(\wh{\theta}_{\lambda^\star}) \le S^2,
\qquad
S^2 = \frac{\Delta^2}{(z_{1-\alpha/2}+z_{1-\beta})^2}.
\]
Therefore
\[
\frac{1}{n}\left(
V_{YY}
- \frac{V_{Yf}^2}{V_{ff}}\cdot\frac{N}{N+n}
\right)\le S^2.
\]
Multiplying through by $n(N+n)>0$ gives
\[
S^2 n^2 + (S^2N - V_{YY})n - N\left(V_{YY}-\frac{V_{Yf}^2}{V_{ff}}\right)\ge 0.
\]
Since this quadratic opens upward, the feasible region is $n$ greater than or
equal to its positive root:
\[
n \ge
\frac{
V_{YY} - S^2N
+
\sqrt{(V_{YY}-S^2N)^2 + 4S^2N\left(V_{YY} - \frac{V_{Yf}^2}{V_{ff}}\right)}
}{2S^2},
\]
which is the fixed-$N$ formula in Proposition~\ref{prop:regression-n}.

Finally, when $N\gg n$, we have $N/(N+n)=1+o(1)$, so
\[
\Var(\wh{\theta}_{\lambda^\star})
\approx
\frac{1}{n}\left(V_{YY}-\frac{V_{Yf}^2}{V_{ff}}\right),
\]
and solving $\Var(\wh{\theta}_{\lambda^\star})\le S^2$ gives
\[
n \ge \frac{V_{YY}-V_{Yf}^2/V_{ff}}{S^2},
\]
which is the large-$N$ simplification stated in the proposition.

\subsection{Additional Simulation Figures}
\label{sec:supp-figures}

This appendix subsection collects the supporting simulation figures
that are referenced but not shown in the main text.  They cover the
two-sample and paired validations from Section~\ref{sec:sim-core},
and the robustness checks from Section~\ref{sec:sim-robustness}.

\paragraph{Two-sample and paired settings.}

\begin{figure}[!ht]
\centering
\includegraphics[width=0.95\linewidth]{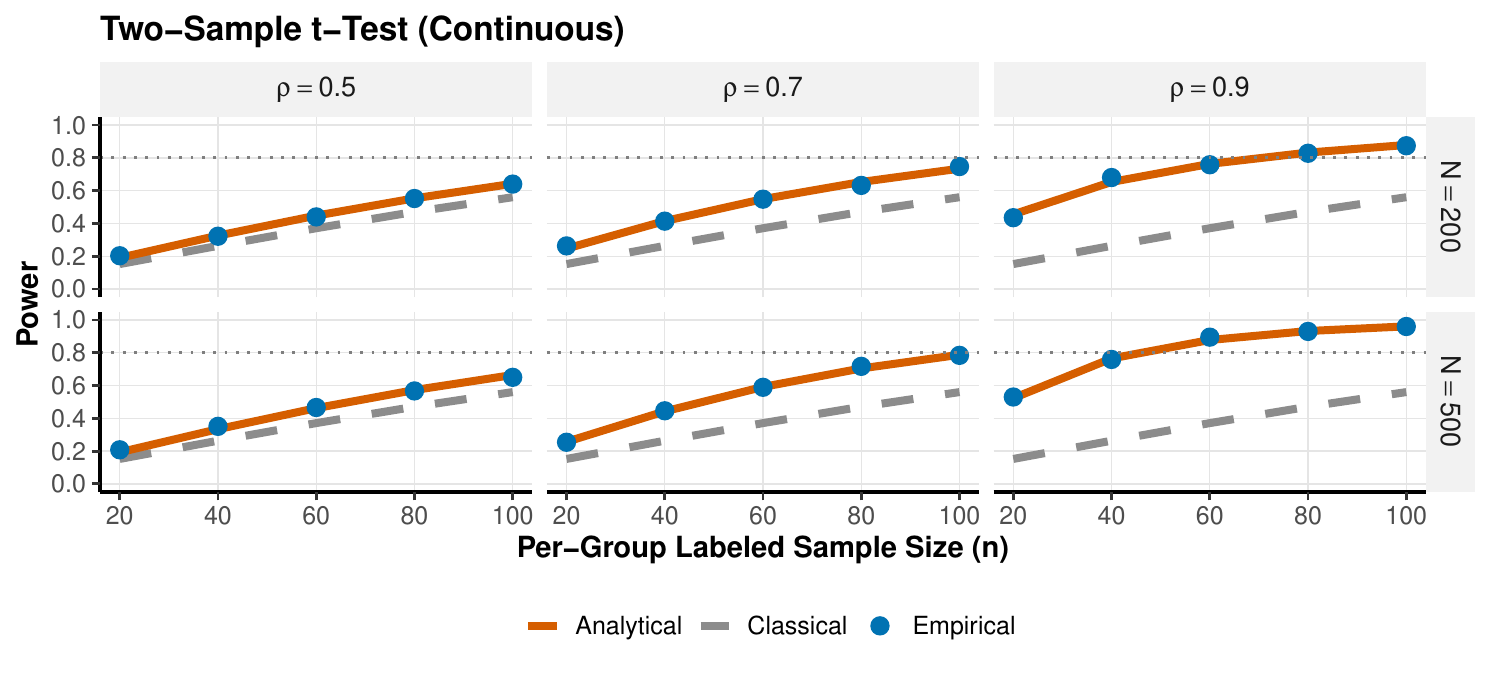}
\caption{Two-sample $t$-test with continuous Gaussian
  outcomes.  Analytical (lines) versus empirical (points) power.}
\label{fig:setting-C}
\end{figure}

\begin{figure}[!ht]
\centering
\includegraphics[width=0.95\linewidth]{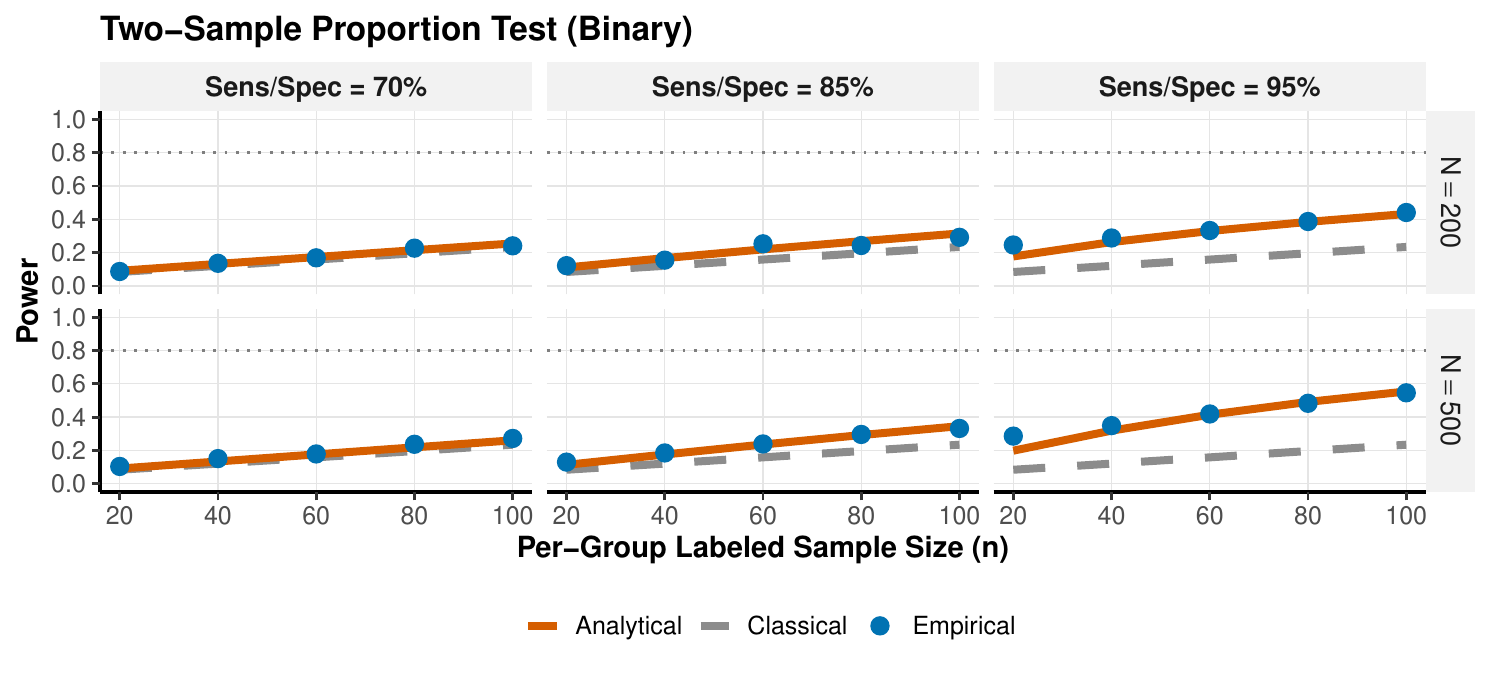}
\caption{Two-sample proportion test with binary outcomes.
  Analytical (lines) versus empirical (points) power.}
\label{fig:setting-D}
\end{figure}

\begin{figure}[!ht]
\centering
\includegraphics[width=0.95\linewidth]{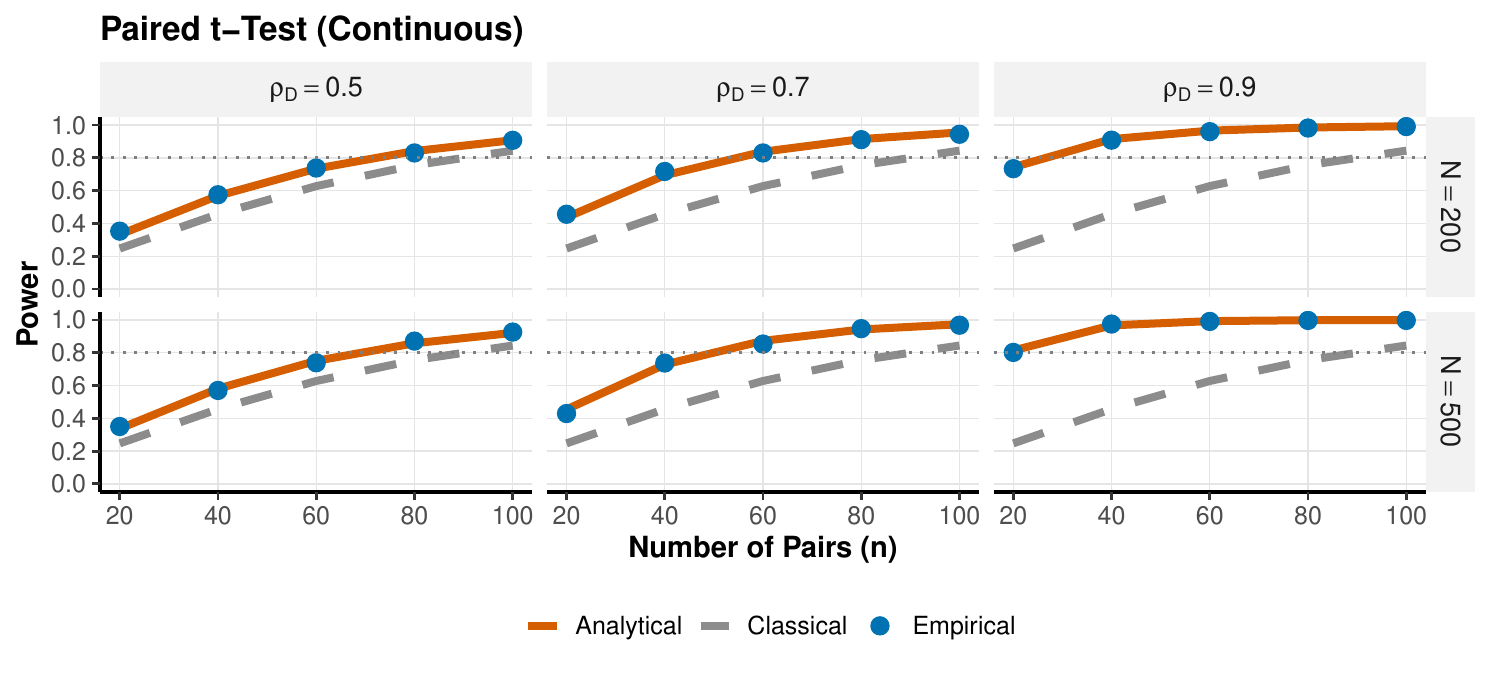}
\caption{Paired $t$-test with continuous differences.
  Analytical (lines) versus empirical (points) power.}
\label{fig:setting-E}
\end{figure}

\begin{figure}[!ht]
\centering
\includegraphics[width=0.95\linewidth]{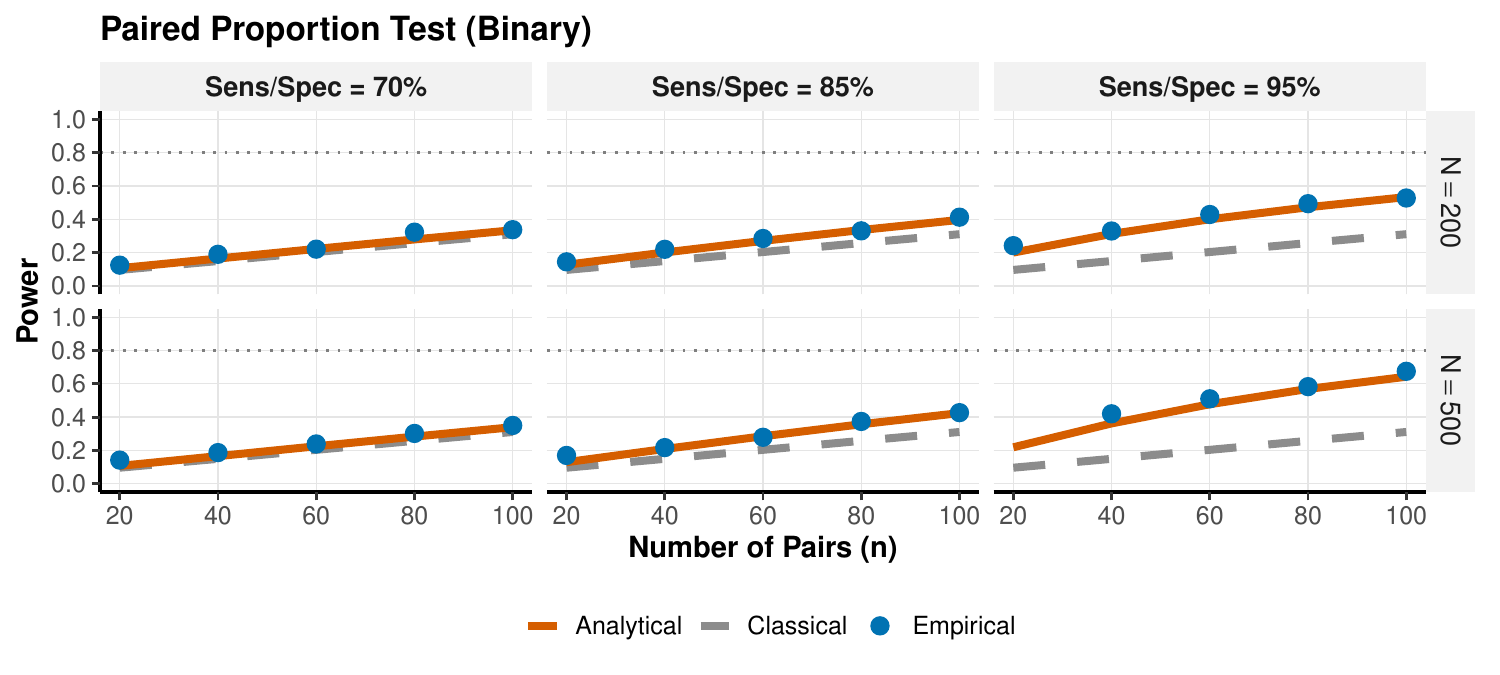}
\caption{Paired proportion test with binary differences.
  Analytical (lines) versus empirical (points) power.}
\label{fig:setting-F}
\end{figure}

\paragraph{Auxiliary planning checks.}

\begin{figure}[!ht]
\centering
\includegraphics[width=0.95\linewidth]{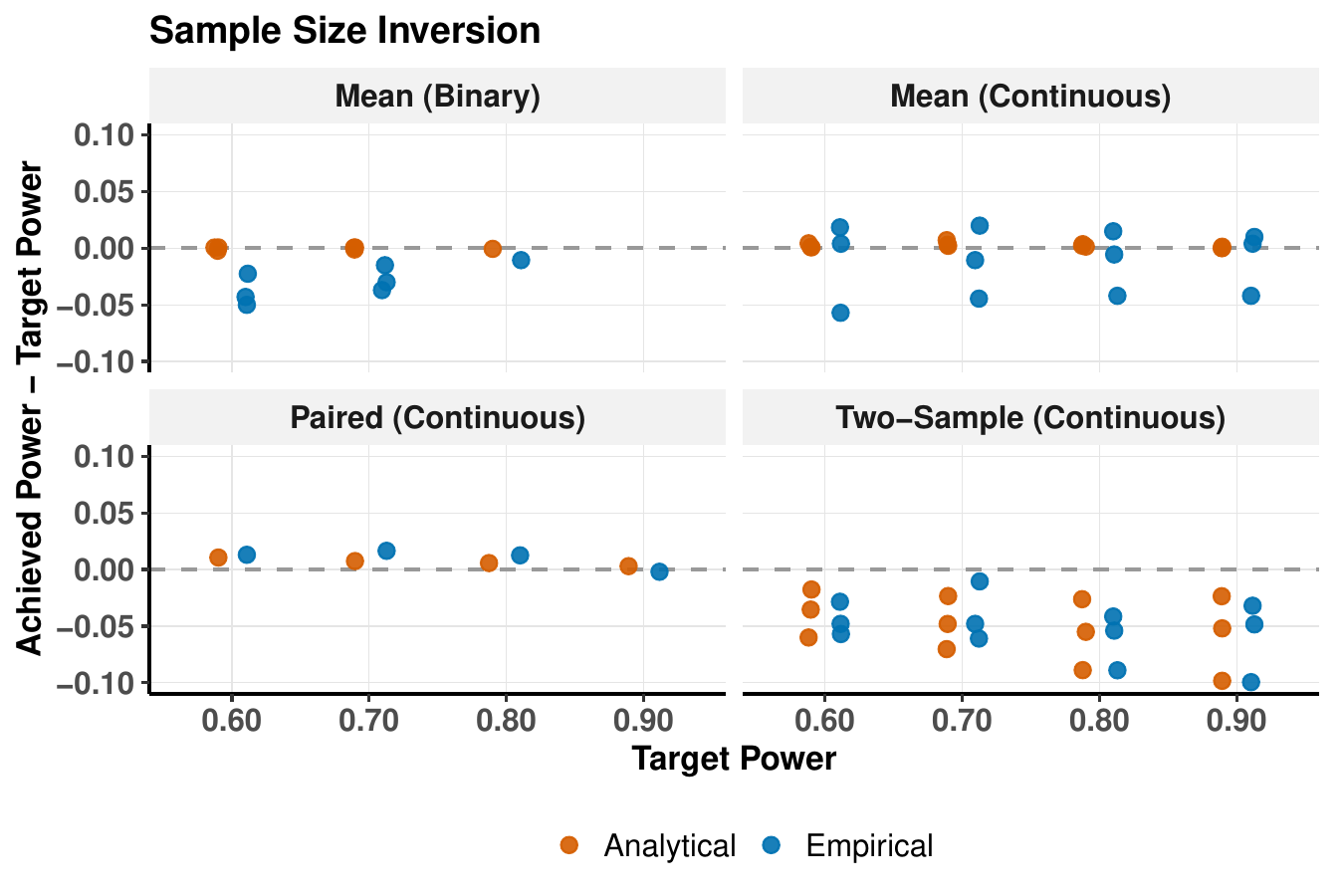}
\caption{Sample-size inversion check across one-sample, two-sample,
  and paired designs.  Points show achieved power minus target power
  at the planned integer $n^\star$; exact inversion would lie on the
  horizontal zero line.}
\label{fig:setting-J}
\end{figure}

\paragraph{Sample-size inversion.}
For continuous one-sample means, binary one-sample means, continuous
two-sample means, and paired continuous designs, we fix target powers
in $\{0.60, 0.70, 0.80, 0.90\}$ and verify that the computed
$n^\star$ achieves power close to the target.  Figure~\ref{fig:setting-J}
plots achieved power minus target power, so exact inversion
corresponds to zero.  Each point is one design configuration within a
test family, such as a particular $\rho$ or binary-surrogate
accuracy, evaluated at the integer $n^\star$ returned by the
planning formula.  The analytical points recompute the asymptotic
power at that integer $n^\star$, while the empirical points are Monte
Carlo rejection rates at the same $n^\star$, so neither series is
forced to equal the target exactly.  Analytical deviations are
uniformly small, while empirical deviations are somewhat larger in the
smallest-$n$ configurations.  Some high-power binary targets at
$N = 500$ required $n^\star > N$ under the fixed-pool convention used
in the planning code and therefore do not appear.

\begin{figure}[!ht]
\centering
\includegraphics[width=0.95\linewidth]{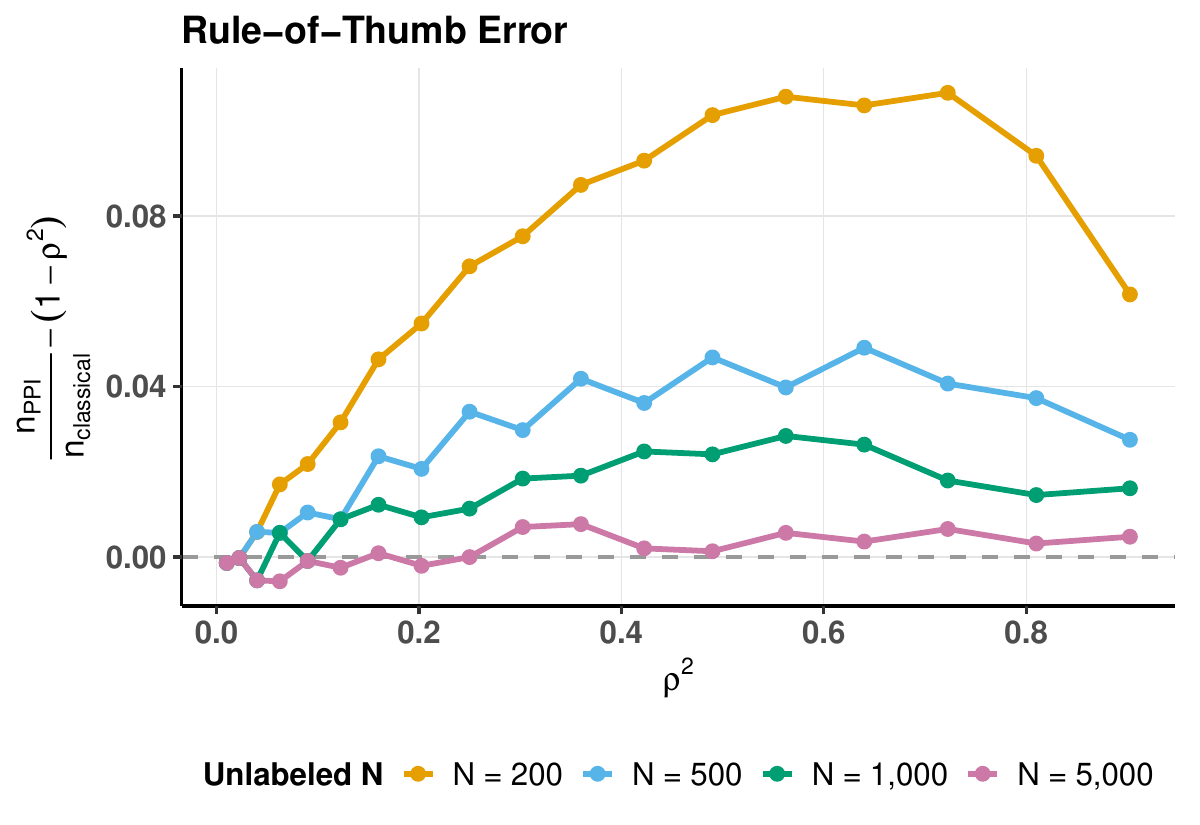}
\caption{Rule-of-thumb error for the labeled-sample reduction ratio.
  Curves plot $n_{\mathrm{PPI}}/n_{\mathrm{cl}} - (1 - \rho^2)$
  against $\rho^2$ for several unlabeled-sample sizes $N$.}
\label{fig:setting-K}
\end{figure}

\paragraph{Rule-of-thumb validation.}
We sweep $\rho \in [0.1, 0.99]$ and
$N \in \{200, 500, 1{,}000, 5{,}000\}$.  Rather than plotting the raw
ratio $n_{\mathrm{PPI}}/n_{\mathrm{cl}}$, Figure~\ref{fig:setting-K}
shows its deviation from the rule-of-thumb approximation
$1 - \rho^2$.  The error shrinks toward zero as $N$ grows, confirming
Corollary~\ref{cor:rule-of-thumb}.

\begin{figure}[!ht]
\centering
\includegraphics[width=0.95\linewidth]{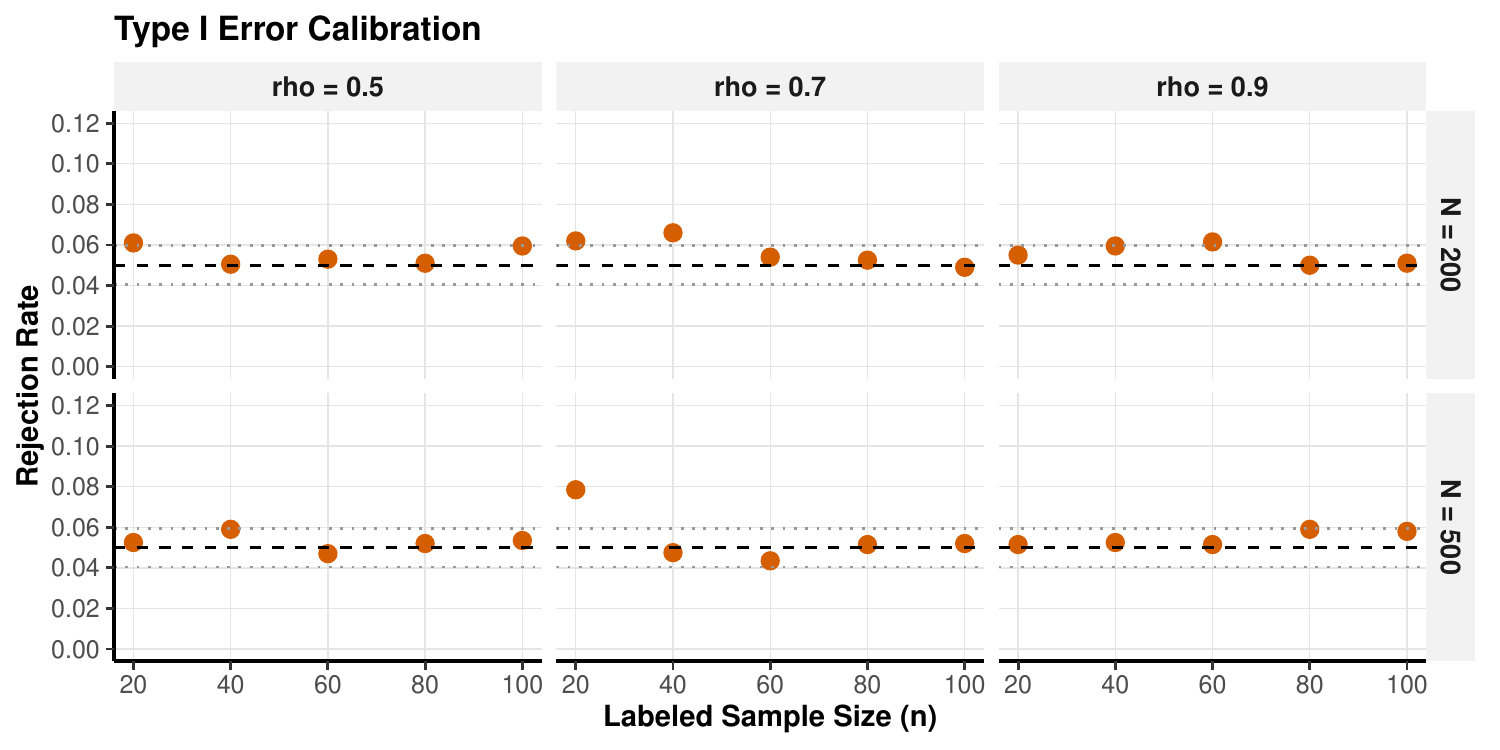}
\caption{Type~I error calibration for the continuous null settings.
  Rejection rates remain close to the nominal level 0.05, with dotted
  lines showing the corresponding Monte Carlo fluctuation band for
  $R = 2{,}000$ replicates.}
\label{fig:setting-L}
\end{figure}

\paragraph{Type~I error calibration.}
We repeat the core one-sample, two-sample, paired, and
distributional-robustness experiments under the null ($\Delta = 0$) with
$R = 2{,}000$ replicates.  Observed \texttt{PPI++} rejection rates are
close to the nominal level: continuous settings range from 0.044 to
0.061, and binary settings from 0.044 to 0.060.  Figure~\ref{fig:setting-L}
shows the continuous null settings; the binary calibration results are
summarized here by their observed range.

\begin{figure}[!ht]
\centering
\includegraphics[width=0.95\linewidth]{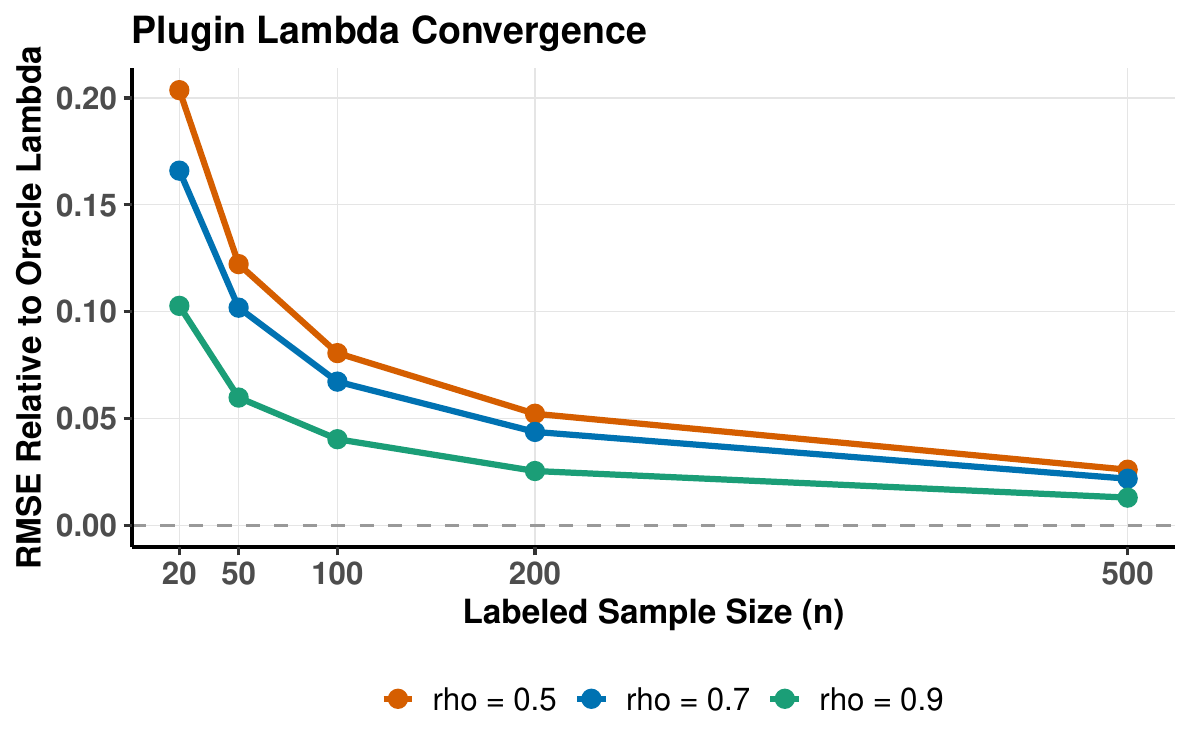}
\caption{Convergence of the plugin $\hat{\lambda}$ toward the oracle
  $\lambda^\star$ as the labeled sample size grows.  The plotted
  quantity is the root-mean-squared error relative to the oracle
  tuning value.}
\label{fig:setting-M}
\end{figure}

\paragraph{Plugin $\lambda$ convergence.}
We verify that the plugin estimate $\hat\lambda$ (computed from
labeled-data sample moments) converges to the oracle $\lambda^\star$
as $n$ increases, with $N = 1{,}000$ fixed.  Figure~\ref{fig:setting-M}
reports the root-mean-squared error of $\hat\lambda$ relative to
$\lambda^\star$, which shrinks toward zero as $n$ grows.

\paragraph{Robustness analyses.}

\begin{figure}[!ht]
\centering
\includegraphics[width=0.95\linewidth]{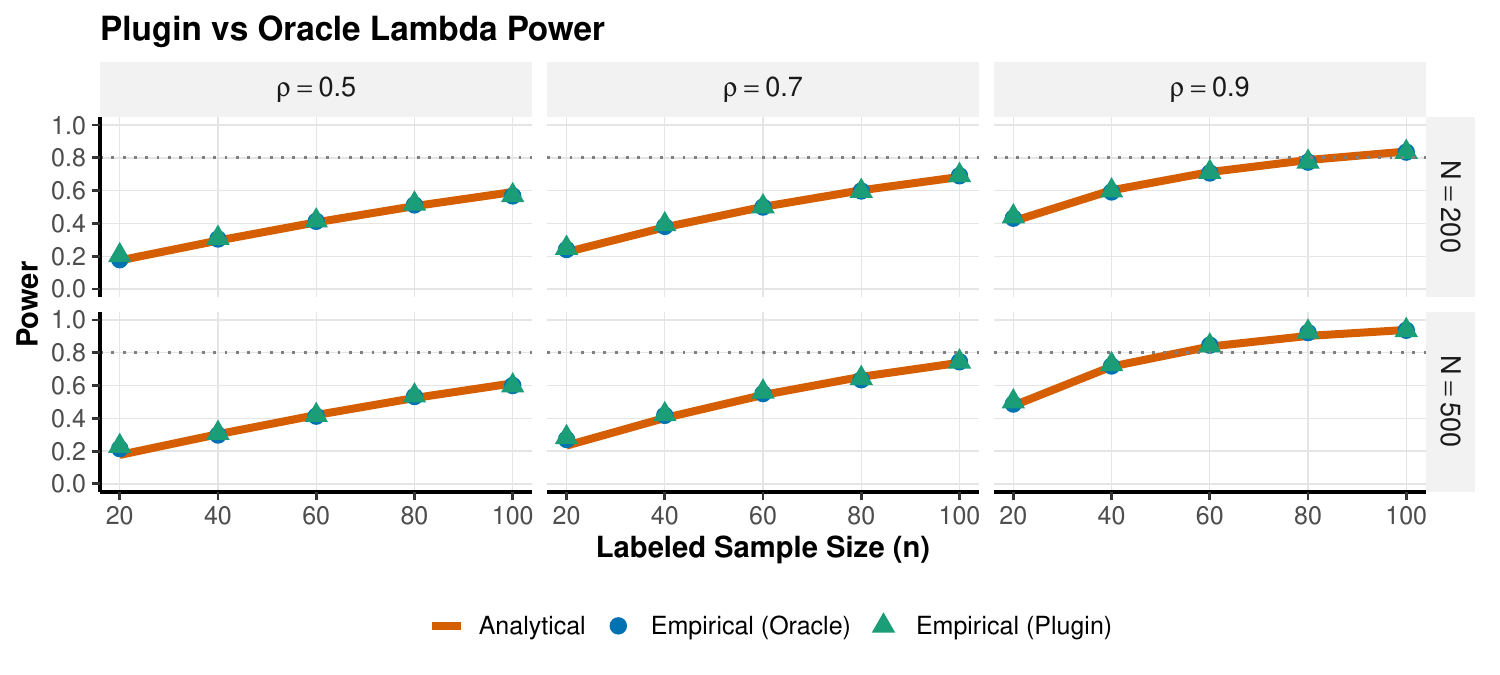}
\caption{Power with plugin versus oracle tuning.  The analytical curve
  uses the oracle variance formula, while the empirical points compare
  oracle and plugin versions of the \texttt{PPI++} test.}
\label{fig:setting-N}
\end{figure}

\paragraph{Plugin versus oracle $\lambda$.}
All previous robustness plots use the oracle $\lambda^\star$ computed
from the true population moments.  In practice, $\lambda$ must be
estimated from data.  This experiment reruns the Gaussian one-sample
DGP from the main validation grid on the same
$n \in \{20, 40, 60, 80, 100\}$ and $N \in \{200, 500\}$ grid, with
both the oracle and a plugin $\hat\lambda$ estimated from labeled-data
covariance.  Figure~\ref{fig:setting-N} shows that empirical power
under the plugin $\hat\lambda$ closely tracks the oracle empirical
curve; the analytical curve (based on the oracle variance) remains a
close approximation even at the smallest design point $n = 20$.
Quantitatively, the maximum discrepancy is 0.023 for oracle
$\lambda^\star$ and 0.037 for plugin $\hat\lambda$, while the
maximum oracle--plugin empirical difference is 0.031.

\begin{figure}[!ht]
\centering
\includegraphics[width=0.95\linewidth]{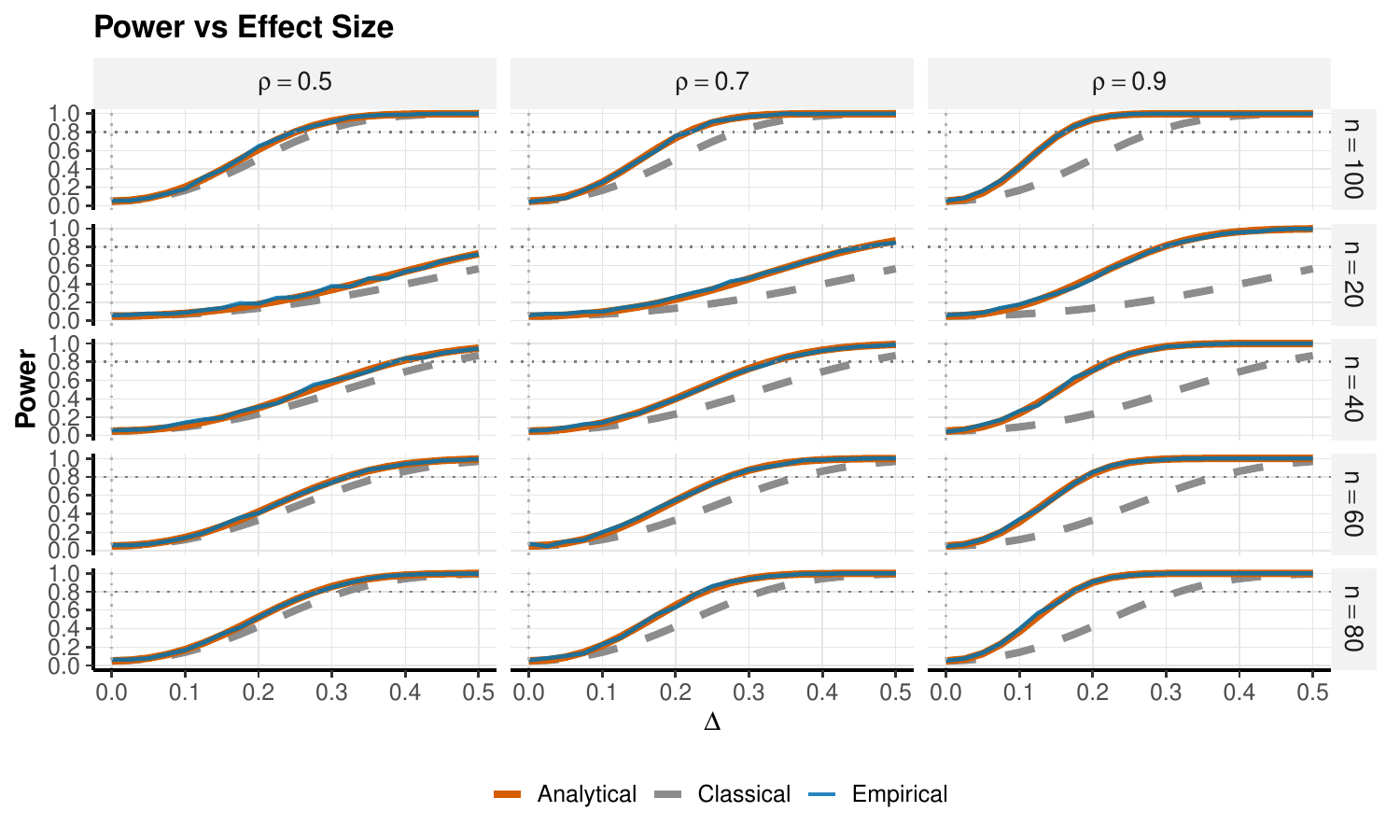}
\caption{Power as a function of effect size for the Gaussian
  one-sample mean problem.  Higher prediction quality shifts the
  \texttt{PPI++} curve leftward relative to the classical design.}
\label{fig:setting-O}
\end{figure}

\paragraph{Power as a function of effect size.}
While the main validation figures fix $\Delta$ and vary $n$, it is also
standard to examine power as a function of effect size at fixed sample
sizes.  Here we sweep $\Delta \in [0, 0.5]$ for
$n \in \{20, 40, 60, 80, 100\}$ with $N = 500$.
Figure~\ref{fig:setting-O} displays the characteristic S-shaped
power curves; higher $\rho$ shifts the curve to the left, requiring
smaller $\Delta$ to achieve any given power level.  The maximum
discrepancy is 0.050.  This setting does not introduce a new variance
formula; it simply visualizes the same one-sample planning problem as
the main one-sample validation figures, but on an effect-size axis
rather than a sample-size axis.

\begin{figure}[!ht]
\centering
\includegraphics[width=0.95\linewidth]{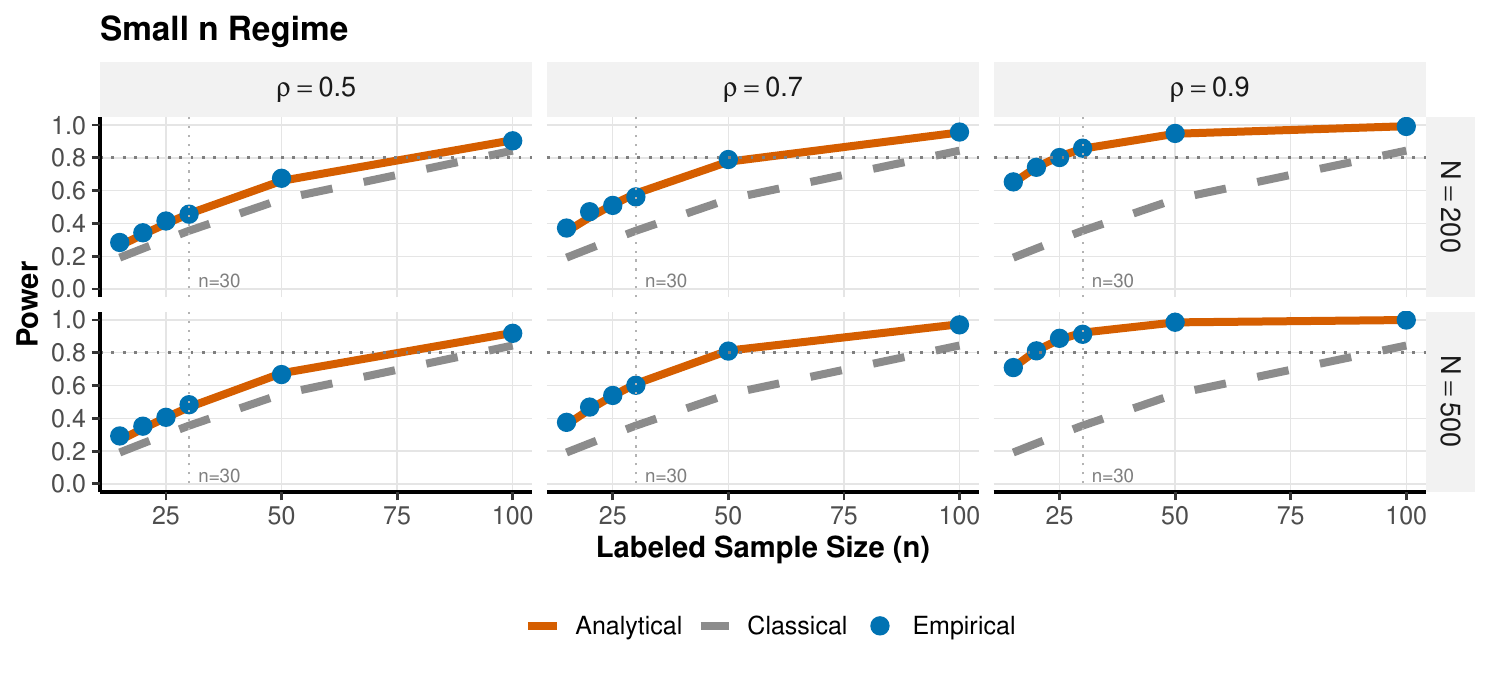}
\caption{Small-$n$ regime for the Gaussian one-sample mean problem.
  Agreement remains usable down to $n = 15$, with the largest
  discrepancies concentrated in the smallest labeled samples.}
\label{fig:setting-P}
\end{figure}

\paragraph{Small labeled samples.}
The power formula relies on the normal approximation, which may be
inaccurate when $n$ is very small.  Here we push to
$n \in \{15, 20, 25, 30, 50, 100\}$ with $R = 2{,}000$ replicates
for tighter Monte Carlo estimates.  Figure~\ref{fig:setting-P}
shows that the analytical formula remains usable for Gaussian
outcomes even at small~$n$.  The maximum discrepancy is 0.036 for
$n \le 25$ and much smaller (0.009) for $n \ge 50$, suggesting that
the normal approximation is most reliable once $n$ is at least
moderate.

\begin{figure}[!ht]
\centering
\includegraphics[width=0.95\linewidth]{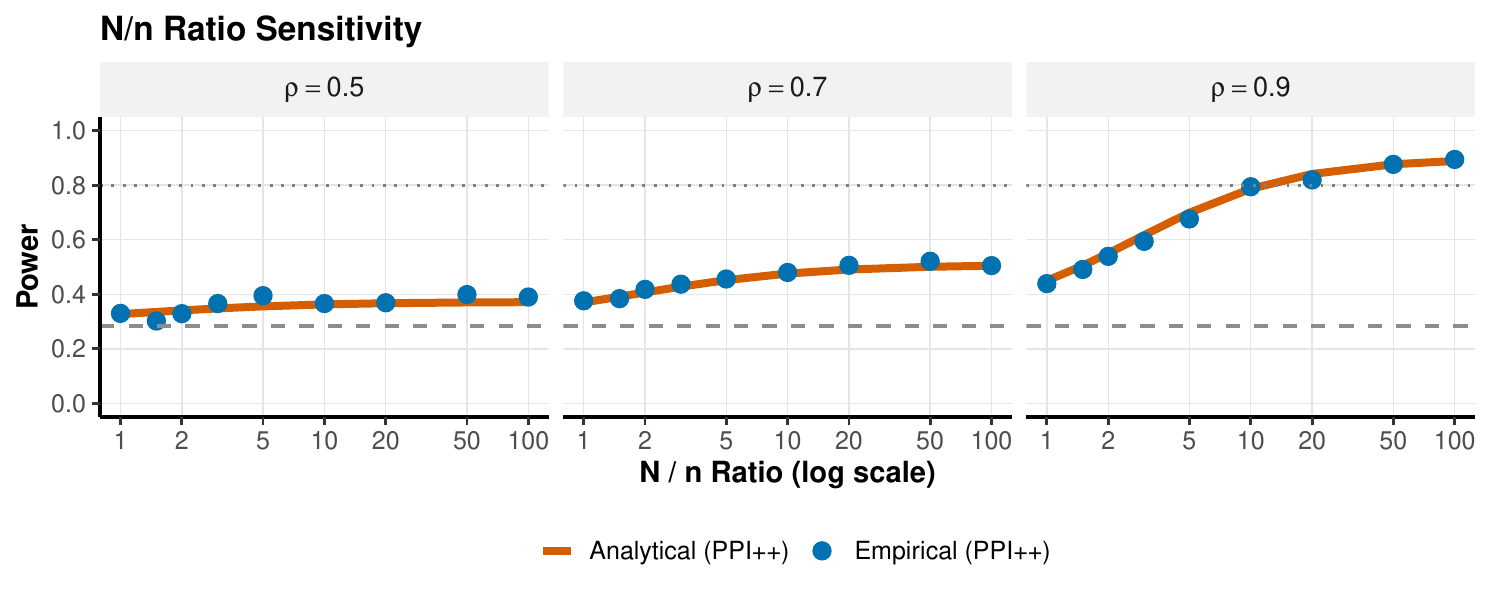}
\caption{Sensitivity to the unlabeled-to-labeled ratio $N/n$.  Power
  gains increase quickly at first and then saturate once the
  unlabeled pool is moderately large.}
\label{fig:setting-Q}
\end{figure}

\paragraph{$N/n$ ratio sensitivity.}
Current PPI practice typically assumes $N \gg n$. Here we fix
$n = 50$ and vary $N/n$ from $1$ to $100$.
Figure~\ref{fig:setting-Q} reveals three findings.  First, at
$N/n = 1$, the \texttt{PPI++} power gain is smaller than in high-ratio
regimes but can still be material when $\rho$ is high.  Second, the
gain saturates: most of the benefit accrues by
$N/n \approx 10$--$20$.  Third, the gain is modulated by $\rho$:
at $\rho = 0.5$, even $N/n = 100$ provides only a modest
improvement, while at $\rho = 0.9$, the improvement is substantial.

\begin{figure}[!ht]
\centering
\includegraphics[width=0.95\linewidth]{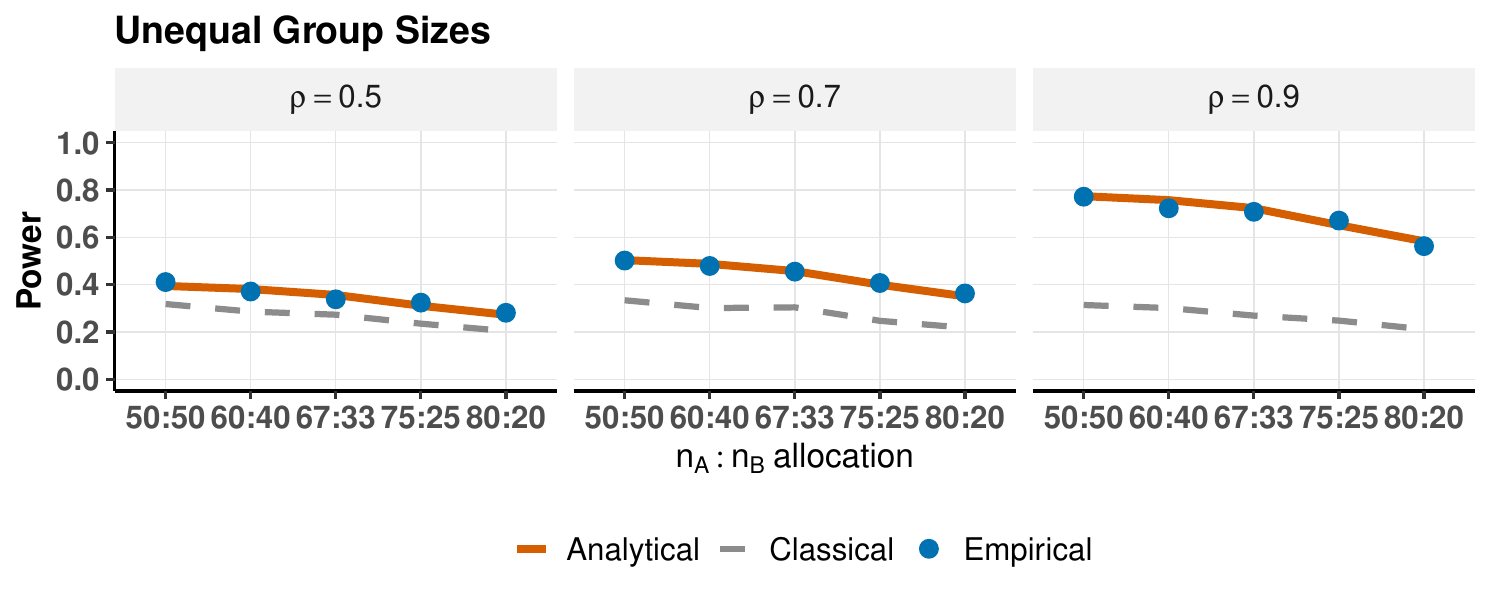}
\caption{Unequal-group two-sample designs.  Power decreases with
  stronger allocation imbalance, while the analytical formula
  continues to track the empirical results closely.}
\label{fig:setting-R}
\end{figure}

\paragraph{Unequal group sizes.}
Two-sample tests in practice often have unbalanced designs.
Here we fix the total labeled budget
($n_A + n_B = 100$, $N_A + N_B = 600$) and vary the allocation
ratio $n_A : n_B \in \{1{:}1, 1.5{:}1, 2{:}1, 3{:}1, 4{:}1\}$,
allocating the unlabeled totals in the same ratio so that
$N_A : N_B = n_A : n_B$.
Figure~\ref{fig:setting-R} shows that the analytical formula
accurately tracks empirical power across all ratios.  Power
decreases with increasing imbalance, consistent with the classical
result that balanced allocation maximizes power for a fixed total
budget.  The maximum discrepancy is 0.030.

\paragraph{Misspecified prediction quality.}
In practice, the user must specify $\rho_{Yf}$ at the planning stage
based on pilot data or published benchmarks, but the true correlation
may differ.  This experiment quantifies this sensitivity: for each
$\rho_{\mathrm{plan}} \in \{0.5, 0.7, 0.9\}$, we compute the
required $n^\star$ assuming $\rho_{\mathrm{plan}}$, then evaluate
the actual power at $n^\star$ under the true
$\rho_{\mathrm{true}} = \rho_{\mathrm{plan}} + \delta_\rho$ with
$\delta_\rho \in \{-0.20, -0.15, \ldots, 0.20\}$, retaining only
feasible values with $\rho_{\mathrm{true}} \in (0.01, 0.99)$, and
$R = 2{,}000$ replicates.  When
$\rho_{\mathrm{true}} < \rho_{\mathrm{plan}}$ (the predictor is
worse than expected), the study is underpowered.  Conversely, when
$\rho_{\mathrm{true}} > \rho_{\mathrm{plan}}$, the study is
conservatively overpowered.  This asymmetry suggests that
practitioners should use conservative (lower-bound) estimates of
prediction quality when planning studies, analogous to the common
advice to use conservative effect-size estimates in classical power
analysis.

\begin{figure}[!ht]
\centering
\includegraphics[width=0.95\linewidth]{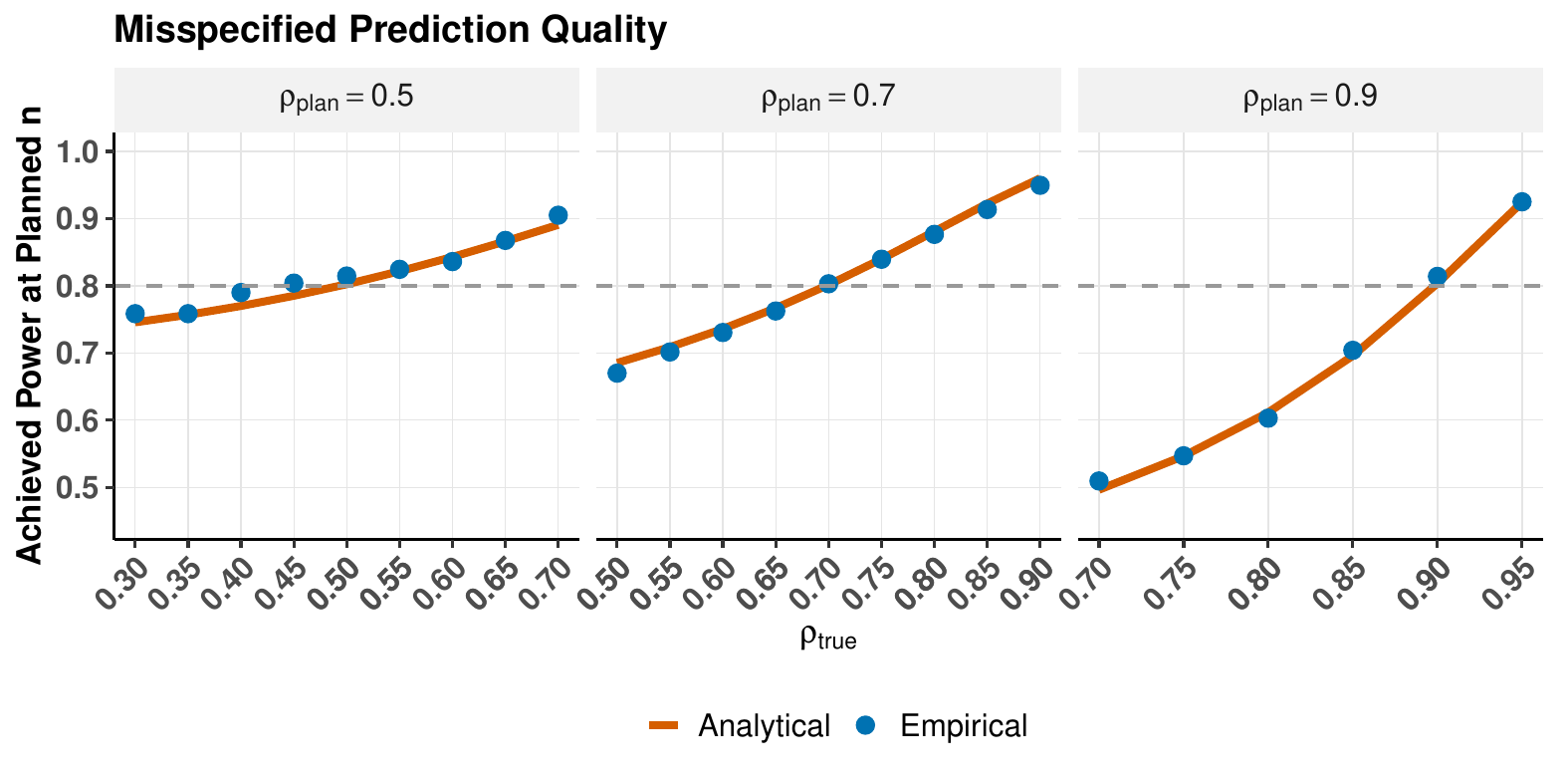}
\caption{Sensitivity to misspecified prediction quality.  The dashed
  line marks the target power, while each panel varies the true
  prediction quality around the value used at the planning stage.}
\label{fig:setting-19}
\end{figure}

\paragraph{Distributional robustness.}

\begin{figure}[!ht]
\centering
\includegraphics[width=0.95\linewidth]{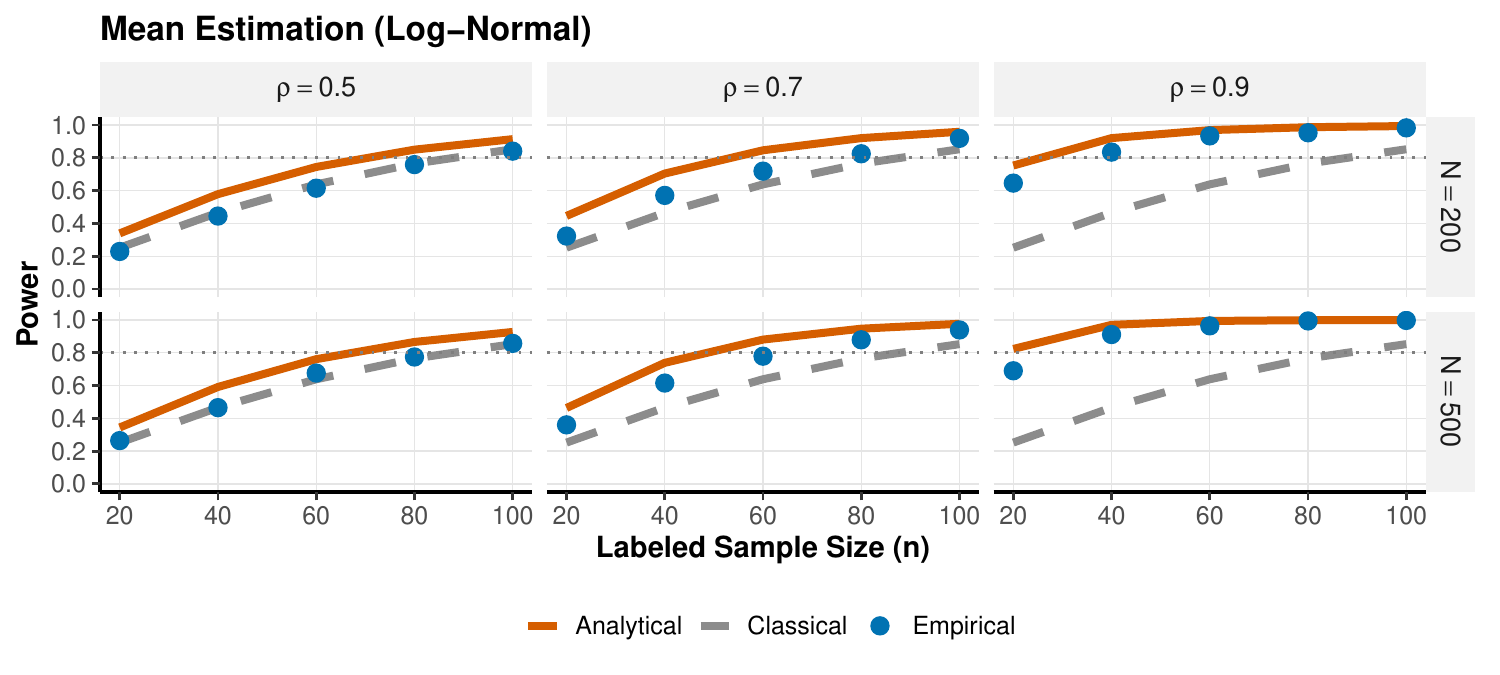}
\vspace{0.4em}
\includegraphics[width=0.95\linewidth]{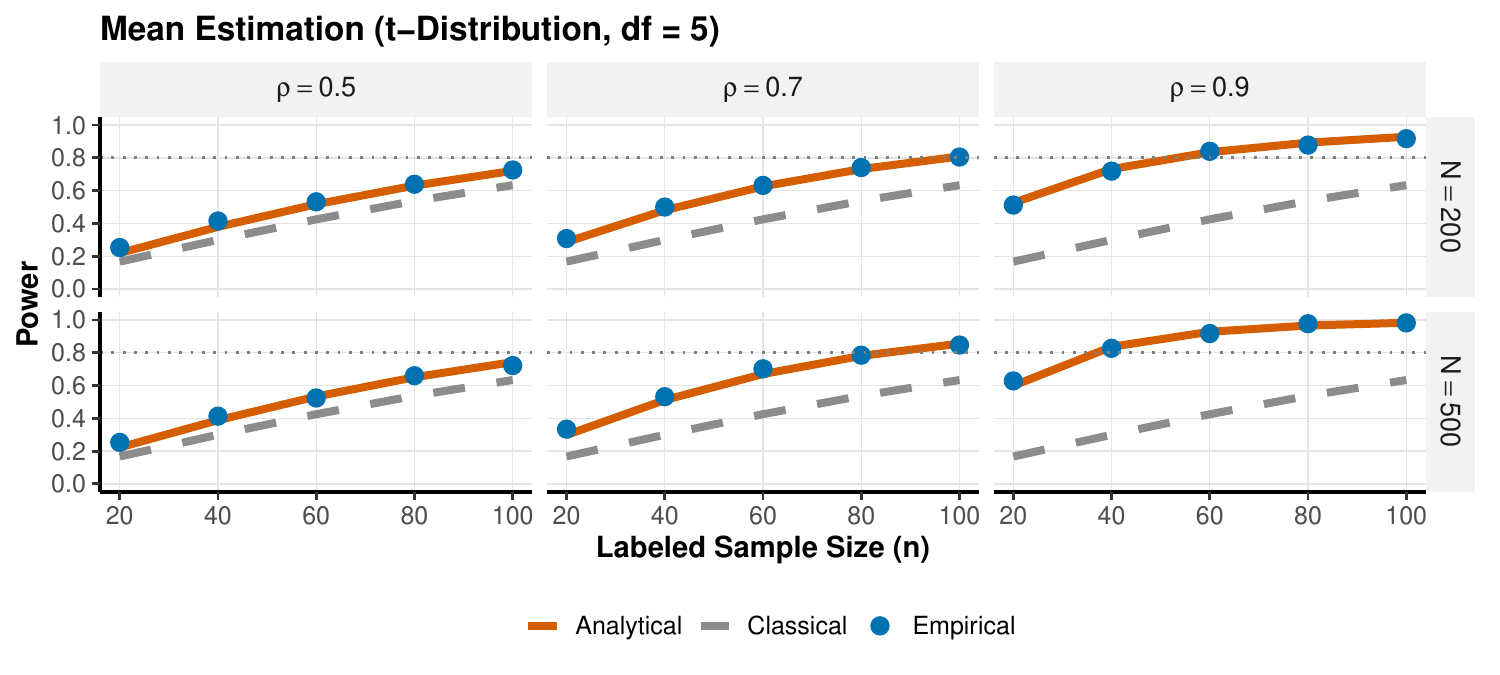}
\caption{Distributional robustness for the one-sample mean problem.
  Top: log-normal outcomes, where the CLT-based formula overshoots at
  small~$n$ under severe skew (max discrepancy 0.144 at $n = 50$,
  $\rho = 0.5$). Bottom: $t_5$ outcomes, where agreement remains tight
  despite heavy tails (max discrepancy 0.018).}
\label{fig:setting-GH}
\end{figure}

Log-normal (right-skewed) and $t_5$ (heavy-tailed) outcomes probe how
far the Gaussian approximation can
be pushed.  For $t_5$ outcomes, agreement remains tight, whereas the
log-normal setting shows larger discrepancies in the most skewed,
small-$n$ regime; see Figure~\ref{fig:setting-GH}.

